\documentclass[a4paper,11pt]{article}
\pdfoutput=1 

\usepackage{jheppub} 

\usepackage[T1]{fontenc} 

\title{\boldmath Hard Photons in Hadroproduction of Top Quarks with Realistic Final
States}

\author[a]{G. Bevilacqua,}
\author[b]{H. B. Hartanto,}
\author[c]{M. Kraus,}
\author[d]{T. Weber }
\author[d]{and M. Worek }

\affiliation[a]{MTA-DE Particle Physics Research Group, University of
Debrecen, H-4010 Debrecen, PBox 105, Hungary} 
\affiliation[b]{Institute for
Particle Physics Phenomenology, Department of Physics, 
Durham University, Durham, DH1 3LE, UK} 
\affiliation[c]{Humboldt-Universit\"at zu
Berlin, Institut f\"ur Physik, Newtonstra\ss{}e 15, D-12489 Berlin,
Germany}
\affiliation[d]{ Institute for Theoretical Particle Physics
and Cosmology, RWTH Aachen University, D-52056 Aachen, Germany}
 
\emailAdd{\\
giuseppe.bevilacqua@science.unideb.hu,
  heribertus.b.hartanto@durham.ac.uk,
  manfred.kraus@physik.hu-berlin.de, \\tweber@physik.rwth-aachen.de,\\
  worek@physik.rwth-aachen.de}

\abstract{ We present a complete description of top quark pair
  production in association with a hard photon in the dilepton
  channel. Our calculation is accurate to NLO in QCD. It is based on
  matrix elements for $e^+\nu_e \mu^-\bar{\nu}_\mu b \bar{b}\gamma$
  production and includes all resonant and non-resonant diagrams,
  interferences, and off-shell effects of the top quarks and the $W$
  gauge bosons.  This calculation constitutes the first full
  computation for top quark pair production with a final state photon
  in hadronic collisions at NLO in QCD. Numerical results for total
  and differential cross sections are presented for the LHC at a
  centre-of-mass energy of $\sqrt{s}=13$ TeV.   For a few
    observables relevant for new physics searches, beyond some
    kinematic bounds, we observe shape distortions of more than
  $100\%$. In addition, we confirm that the size of the top quark
  off-shell effects for the total cross section is consistent with the
  expected uncertainties of the narrow width approximation. Results
  presented here are not only relevant for beyond the Standard Model
  physics searches but also important for precise measurements of the
  top-quark fiducial cross sections and top-quark properties at the
  LHC.}

\dedicated{\rm TTK-18-09, HU-EP-18/07, IPPP/18/17}

\keywords{NLO Computations, QCD Phenomenology, Heavy Quark Physics}

\begin{document} 
\maketitle
\flushbottom

\tableofcontents

%
\section{Introduction}
%

The top quark, discovered by the CDF and D0 experiments at Fermilab
more than 20 years after its existence was postulated to explain the
observed CP violations in kaon decays, is the heaviest elementary
particle in the Standard Model (SM) of particle physics. Due to its
large mass and its correspondingly short lifetime the top quark decays
before hadronic bound states can be formed, thus, passing its spin
information onto its decay products. With a mass of the order of the
electroweak scale, the top quark Yukawa coupling to the Higgs boson is
of the order of unity. This alone makes the top quark unique among the
fermions and its potential to provide insights into physics beyond the
SM (BSM) is anticipated. Various BSM models introduce modifications
within the top quark sector, which can be constrained by precise
measurements of the $t\bar{t}$ and ${t\bar{t}+X}$ cross sections,
 where $X=H,j,\gamma,Z,W^\pm, t\bar{t}$. Examples include
composite top quarks, Randall-Sundrum extra dimensions, models with
coloured scalars or universal extra dimensions.  Studies of top quark
properties provide a unique environment for testing the SM and for
hunting BSM physics.  Investigations of the dynamics of the top quark
pair production process in association with a hard photon, for
example, directly probe the top quark electric charge and the
structure of the $t\bar{t}\gamma$ coupling. Any deviation from the SM
prediction of the measured observables could be an indication of BSM
physics and might be linked to the production of an exotic (possibly
heavier) top-like quark or the top quark with an anomalous electric
dipole moment, see e.g. Refs.
\cite{Fael:2013ira,Saavedra:2014vta,Schulze:2016qas,Etesami:2016rwu}.
Good examples of such observables comprise the transverse momentum
spectrum of the photon, $p_{T,\gamma}$, and the azimuthal
angle-rapidity distance between the photon and the softest $b$-jet,
$\Delta R_{b_2,\gamma}$, \cite{Baur:2001si,Baur:2004uw}.

First evidence for $t\bar{t}\gamma$ production has been reported by
the CDF collaboration in $p\bar{p}$ collisions at the Tevatron with
$\sqrt{s}=1.96$ TeV \cite{Aaltonen:2011sp}. Observation was also
announced by the ATLAS collaboration in $pp$ collisions at the Large
Hadron Collider (LHC) with $\sqrt{s}=7$ TeV
\cite{Aad:2015uwa}. Meanwhile, measurements have been carried out at
the LHC by both ATLAS and CMS collaborations at $\sqrt{s}=8$ TeV
\cite{Aaboud:2017era,Sirunyan:2017iyh}. For now, due to small
available statistics, these measurements only comprise cross
sections. However, with the second run of the LHC at $\sqrt{s}=13$ TeV
and with increased luminosity more exclusive observables and various
properties of the top quark can be scrutinised.

On the theory side, first next-to-leading order (NLO) QCD calculations
for $t\bar{t}\gamma$ have been performed for on-shell top quarks
\cite{PengFei:2009ph, PengFei:2011qg,Maltoni:2015ena}. Recently, even
NLO electroweak corrections have been completed
\cite{Duan:2016qlc}. Computations in the approximation of stable tops
give a general idea about the size of higher order effects. However,
they can not provide a reliable description of top quark decay
products or the magnitude of NLO corrections when specific cuts are
imposed on the final states. For a realistic analysis, not only higher
order effects to $t\bar{t}\gamma$ production need to be included but
also radiative decays of top quarks have to be incorporated. This has
been (partially) achieved by means of parton showers through matching
fixed order NLO QCD predictions for $t\bar{t}\gamma$ with parton
shower programs via the \textsc{Powheg} method \cite{Nason:2004rx,
Frixione:2007vw}, albeit omitting photon emissions in the parton
shower evolution and $t\bar{t}$ spin correlations
\cite{Kardos:2014zba}. A more sophisticated approach has been employed
in \cite{Melnikov:2011ta}, where NLO QCD corrections to production and
decays in the so called narrow width approximation (NWA) have been
calculated.  Non-factorisable QCD contributions, however, that imply a
cross talk between production and decays of top quarks and which
require going beyond the NWA, have been so far neglected.  Such
contributions are formally suppressed, i.e.  ${\cal O}(\Gamma_t/m_t)
\approx 0.8\%$, where $\Gamma_t\,,m_t$ are the top quark width and
mass respectively. They proved to be small in the inclusive cross
section. Nonetheless, they can be strongly enhanced in case of
exclusive observables that are crucial for new physics searches. The
lack of any evidence of BSM at the LHC has put known new physics
scenarios under significant strain.  Our attention is focused now on
precision physics and indirect searches aiming at deviations from SM
predictions in precision observables. To probe more subtle BSM effects
also in $t\bar{t}\gamma$ production, state of the art theoretical
predictions for this process are of vital importance.

In this paper, we calculate for the first time the NLO QCD corrections
to the fully realistic final state $pp \to e^+ \nu_e \mu^-
\bar{\nu}_\mu b\bar{b} \gamma$. We consistently take into account
resonant and non-resonant top quark and $W$ gauge boson contributions
and interference effects among them. Our theoretical predictions are
presented in the form of the fully flexible Monte Carlo program. Thus,
various observables and cuts can be explored and their usefulness can
be demonstrated in realistic Monte Carlo simulations.  The
final results are provided as the Ntuple files
\cite{Bern:2013zja}. Specifically,  they are stored in the form of the
modified Les Houches event files \cite{Alwall:2006yp} and ROOT files
\cite{Antcheva:2009zz} that might be directly employed in experimental studies at
the LHC.

As a final comment, we note that
NLO QCD corrections with complete top quark off-shell effects are also
known for $t\bar{t}$, $t\bar{t}H$ and $t{\bar t}j$ productions
\cite{Denner:2010jp,Bevilacqua:2010qb,Denner:2012yc,
Frederix:2013gra,Heinrich:2013qaa,Denner:2017kzu,Denner:2015yca,
Bevilacqua:2015qha,Bevilacqua:2016jfk}. In case of $t\bar{t}$ and
$t\bar{t}H$ NLO electroweak corrections have been added as well
\cite{Denner:2016jyo,Denner:2016wet}.

The article is organised as follows. In Section \ref{sec:2} we
describe the details of our calculation. Input parameters and cuts to
simulate detector response are summarised in Section
\ref{sec:3}. Numerical results for the integrated and differential
cross sections for the LHC Run II energy of $13$ TeV for two
renormalisation ($\mu_R$) and factorisation ($\mu_F$) scale choices
are presented in Sections \ref{sec:4} and \ref{sec:5}, respectively.  
The theoretical uncertainties of the total cross sections and various
differential cross sections, that are associated with neglected higher
order terms in the perturbative expansion and with  different
parametrisations of the parton distribution functions (PDFs), are also
given there.  Finally, in Section \ref{summary} our conclusions are
given.

%
\section{Computational Framework}
\label{sec:2}
%

\begin{figure*}[t!]
\begin{center}
\includegraphics[width=1.0\textwidth]{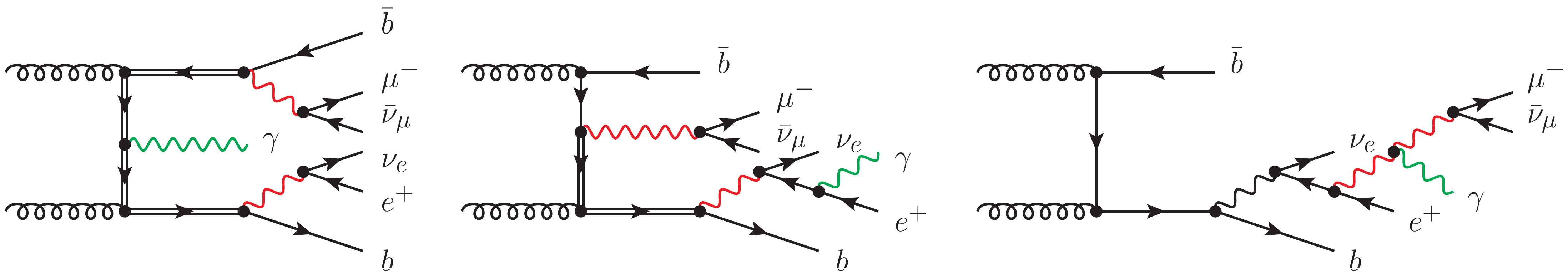}
\end{center}
\vspace{-0.6cm}
\caption{\it Representative Feynman diagrams, involving two (first
  diagram), one (second diagram) and no top quark resonances (third
  diagram),  contributing to   $pp \to e^+\nu_e \mu^-
  \bar{\nu}_\mu b\bar{b}\gamma$ production at leading order.}
\label{fig:fd}
\end{figure*}

At leading order (LO) in the perturbative
expansion, the $pp\to e^+ \nu_e \mu^- \bar{\nu}_\mu b\bar{b} \gamma$
final state is produced via the scattering of either two gluons or a quark and
the corresponding anti-quark
\begin{equation}
\begin{split}
gg&\to e^+ \nu_e \mu^- \bar{\nu}_\mu b\bar{b} \gamma\,,\\[0.2cm]
q\bar{q}&\to e^+ \nu_e \mu^- \bar{\nu}_\mu b\bar{b} \gamma\,,
\end{split}
\end{equation}
where $q$ stands for $u,d,c,s$. In total, the $gg\to e^+ \nu_e \mu^-
\bar{\nu}_\mu b\bar{b} \gamma$ subprocess comprises 628 Feynman
diagrams and the $q\bar{q}\to e^+ \nu_e \mu^- \bar{\nu}_\mu b\bar{b}
\gamma$ subprocess has only 346 tree level Feynman diagrams. Even
though we do not use Feynman diagrams to obtain matrix elements for
each subprocess, we provide such information here 
in order to shed some lights on the
complexity of the process at hand.  A few examples of Feynman
diagrams contributing at ${\cal O}(\alpha_s^2\alpha^5)$ for the $gg$
initial state are presented in Fig.~\ref{fig:fd}.

The calculation of scattering amplitudes is performed by means of an
automatic off-shell iterative algorithm \cite{Papadopoulos:2005ky},
which is implemented within the \textsc{Helac-Dipoles} package
\cite{Czakon:2009ss} and the \textsc{Helac-Phegas} Monte Carlo (MC)
program \cite{Cafarella:2007pc}. The latter framework has been used to
cross check our LO results. For the phase-space integration depending
on the MC framework \textsc{Phegas} \cite{Papadopoulos:2000tt},
\textsc{Parni} \cite{vanHameren:2007pt} and \textsc{Kaleu}
\cite{vanHameren:2010gg} have been employed.
%
%

At NLO, virtual corrections are obtained from the interference of the
one-loop dia- grams with the tree level amplitude. They might be
classified into self-energy, vertex-, box-, pentagon-, hexagon- and
heptagon-type corrections. For the $gg$ dominant production channel we
have $36032$ Feynman diagrams at one-loop, among these $90$ are
heptagons and $958$ are hexagons.  The latter numbers have been
obtained with the help of \textsc{Qgraf} \cite{Nogueira:1991ex}.
Virtual corrections are evaluated in $d = 4 - 2\epsilon$ dimensions in
the 't Hooft-Veltman version of the dimensional regularisation and
using the Feynman gauge for gauge bosons. The singularities coming
from infrared divergent pieces are canceled by the corresponding ones
arising from the counter-terms of the adopted subtraction scheme
integrated over the phase space of the unresolved parton. The finite
contributions of the loop diagrams are evaluated numerically in $d =
4$ dimensions. To ensure numerical stability of our calculations we
have used the Ward identity test. On-shell transversality of gluon
amplitudes has been checked up to the one loop level for every phase
space point. About $6\%$ of events, that fail the gauge-invariance
check, have been recomputed in quadruple precision. For
  $q\bar{q}\to e^+ \nu_e \mu^- \bar{\nu}_\mu b\bar{b} \gamma$ partonic
  subprocess at ${\cal O}(\alpha_s^3\alpha^5)$ there are no gluons as external
particles. Since unstable electroweak bosons are treated in the
fixed-width scheme, the photon Ward identity test could not be applied
straightforwardly.  Instead the scale test, as introduced in
Ref.~\cite{Badger:2010nx}, has been performed. It is based on the
simple observation that the momenta can be rescaled and the amplitude
can be recalculated and compared to the original one.  In this case
higher precision has also been used to recompute $0.15\%$ of events
that did not pass the test. Another cross check that we have performed
comprises a verification of the cancelation of infrared poles. We
compute the virtual corrections using \textsc{Helac-1Loop}
\cite{vanHameren:2009dr} and \textsc{CutTools} \cite{Ossola:2007ax},
which are both parts of the \textsc{Helac-NLO} Monte Carlo framework
\cite{Bevilacqua:2011xh}.  The \textsc{CutTools} code contains an
implementation of the OPP method for the reduction of one-loop
amplitudes at the integrand level \cite{Ossola:2006us, Ossola:2008xq,
Draggiotis:2009yb}. For unstable top quarks the complex mass scheme,
as described in Refs.  \cite{Denner:1999gp,Denner:2005fg}, is used.
At the one loop level the appearance of $\Gamma_t\ne 0$ in the
propagator requires the evaluation of scalar integrals with complex
masses, which is supported by the \textsc{OneLOop} program
\cite{vanHameren:2010cp} employed in our studies. For consistency,
mass renormalisation for the top quark is also done by applying the
complex mass scheme in the well known on-shell scheme.  Re-weighting
techniques, helicity and colour sampling methods are employed in order
to optimise the performance of our system.

The real emission corrections to the LO process arise from tree-level
amplitudes with one additional parton, i.e. an additional gluon, or a
quark anti-quark pair replacing a gluon.  All possible subprocesses
contributing to the real emission part are shown in Table
\ref{tab:real}. The number of Feynman diagrams corresponding to the
subprocesses under scrutiny is also given to underline the complexity
of the calculations. The following three subprocesses $qg\to e^+
\nu_e \mu^- \bar{\nu}_\mu b\bar{b} \gamma q$, $\bar{q}g\to e^+ \nu_e
\mu^- \bar{\nu}_\mu b\bar{b} \gamma \bar{q}$ and $q\bar{q}\to e^+
\nu_e \mu^- \bar{\nu}_\mu b\bar{b} \gamma g$ are related by crossing
symmetry.
%
\begin{table}[h!]
\begin{center}
\begin{tabular}{|c|c|c|c|}
\hline
\textsc{Partonic}& \textsc{Number Of} &\textsc{Number Of}&
                                                           \textsc{Number
                                                           Of}\\[0.2cm]
\textsc{Subprocess} &\textsc{Feynman Diagrams} & \textsc{CS Dipoles}  
& \textsc{NS Subtractions} \\[0.2cm]
\hline
$gg\to e^+ \nu_e \mu^- \bar{\nu}_\mu b\bar{b} \gamma g$
&4348 & 27 & 9 \\[0.2cm]
$qg\to e^+ \nu_e \mu^- \bar{\nu}_\mu b\bar{b} \gamma q$ 
& 2344&15&5\\[0.2cm]
$\bar{q}g\to e^+ \nu_e \mu^- \bar{\nu}_\mu b\bar{b} \gamma \bar{q}$ 
& 2344 &15&5\\[0.2cm]
$q\bar{q}\to e^+ \nu_e \mu^- \bar{\nu}_\mu b\bar{b} \gamma g$
&2344 &15 & 5\\[0.2cm]
\hline
 \end{tabular}
\end{center}
\caption{\label{tab:real} \it The list of partonic subprocesses
  contributing to the subtracted real emission for the $pp\to  e^+
  \nu_e \mu^- \bar{\nu}_\mu b\bar{b} \gamma +X$ process. Also shown
  are the number of Feynman diagrams, as well as the number of
  Catani-Seymour and Nagy-Soper subtraction terms. }
 \end{table}
%
The singularities from soft and/or collinear parton emissions are isolated
via subtraction methods for NLO QCD calculations: the commonly used
Catani-Seymour dipole subtraction
\cite{Catani:1996vz,Catani:2002hc,Czakon:2009ss}, and a fairly new
Nagy-Soper subtraction scheme \cite{Bevilacqua:2013iha}, both
implemented in the \textsc{Helac-Dipoles} software. Specifically,
\textsc{Helac-Dipoles} implements the massless dipole formalism of
Catani and Seymour, as well as its massive version for arbitrary
helicity eigenstates and colour configurations of the external
partons. The Nagy-Soper subtraction scheme makes use of random
polarisation and colour sampling of the external partons. An overall
performance of this scheme has been assessed in
Ref.~\cite{Bevilacqua:2013iha} where a detailed comparison with
results based on the Catani-Seymour dipole subtraction scheme has been
carried out for a collection of  processes.  
Thus, in Table \ref{tab:real} we only compare the total number of
subtraction terms that need to be evaluated in both schemes.  In each
case, three times fewer terms are needed in the Nagy-Soper subtraction
scheme compared to the Catani-Seymour scheme. The difference
corresponds to the total number of possible spectators, which are
relevant in the Catani-Seymour case, but not in the Nagy-Soper case. A
phase space restriction ($\alpha_{max}$) on the contribution of the
subtraction terms, as proposed e.g. in Refs.~\cite{Frixione:1995ms,Nagy:1998bb,
Nagy:2003tz,Campbell:2005bb,Bevilacqua:2009zn,Czakon:2015cla}, is
included for both subtraction cases.  We consider two choices, namely
$\alpha_{max}=1$, that corresponds to the original formulation of the
Catani-Seymour and Nagy-Soper subtraction scheme, as well as
$\alpha_{max}=0.01$. In case of the Nagy-Soper subtraction scheme,
which was our default scheme used for the calculations, we have
checked that the final results for the sum of real radiation and
integrated dipoles were independent of the $\alpha_{max}$ choice. We
have further cross checked that results for the real emission part are
in agreement with results obtained with the Catani-Seymour dipole
subtraction scheme.  For the real correction part, we also adopt the
\textsc{Kaleu} phase-space generator that is equipped with additional,
special channels that proved to be important for phase-space
optimisation.

%
\section{Input Parameters and Cuts}
\label{sec:3}
%

In the following we present predictions for $pp \to e^+ \nu_e \, \mu^-
\bar{\nu}_\mu \, b \bar{b} \,\gamma +X$ production at ${\cal
O}(\alpha_s^3 \alpha^5)$ for the LHC Run II energy of ${\sqrt{s} =
13}$ TeV. We consider decays of weak bosons to different lepton
generations only, thus, neglecting the interference effects. However,
the difference between the LO cross sections for $pp \to e^+ \nu_e \, \mu^-
\bar{\nu}_\mu \, b \bar{b} \,\gamma +X$ and $pp \to e^+ \nu_e \, e^-
\bar{\nu}_e \, b \bar{b} \,\gamma +X$ is at the per mille level,
thus, the simplification is very well motivated. The complete cross
section for $pp\to \ell^+ \nu_\ell \ell^-\bar{\nu}_\ell
b\bar{b}\gamma$, with $\ell^\pm = e^\pm,\mu^\pm$, can be obtained by
multiplying results presented in the following by a factor of $4$. The
Cabibbo-Kobayashi-Maskawa mixing of the quark generations is neglected
and the following SM parameters are used
%
\begin{equation}
\begin{array}{lll}
  m_{W}=80.385 ~\text{GeV},
&\quad \quad \quad\quad \quad \quad &\Gamma_{W} = 2.0988 ~\text{GeV},
\vspace{0.2cm}\\
 m_{Z}=91.1876  ~\text{GeV} \,, 
&\quad \quad \quad\quad \quad \quad&\Gamma_{Z} = 2.50782 ~\text{GeV},
\vspace{0.2cm}\\
 \Gamma_{t}^{\rm LO} = 1.47848 ~\text{GeV}\,, &\quad 
\quad \quad\quad \quad \quad&
 \Gamma_{t}^{\rm NLO} = 1.35159  ~\text{GeV}\,,
\vspace{0.2cm}\\
m_{t}=173.2 ~\text{GeV}\,, &\quad \quad\quad \quad \quad \quad&
G_\mu=1.166378 \times 10^{-5}
~\text{GeV}^{-2}\,.
\end{array}
\end{equation}
%
The top quark width is calculated according to \cite{Jezabek:1988iv}
 at the scale $m_t$. All other quarks, including $b$
quarks as well as leptons, are assumed to be massless.  Leptonic $W$
gauge boson decays do not receive NLO QCD corrections. To include some
effects of higher order corrections for the gauge boson widths, the
NLO QCD values of the corresponding $W$ width are used for LO and NLO
matrix elements.  The electroweak coupling is derived from the Fermi
constant $G_\mu$ in the $G_\mu-$scheme, where 
$\alpha_{G_\mu}=\sqrt{2} G_\mu m_W^2 \sin_W^2/\pi$ and
$\sin_W^2=1-m_W^2/m_Z^2$.  For our setup we have 
$\alpha_{G_\mu} \approx 1/132$.  In the $G_\mu$-scheme electroweak
corrections related to the running of $\alpha_{G_\mu}$ and to the $\rho$
parameter (proportional to $m_t^2/m^2_W$), are incorporated. By
parametrising the lowest order in terms of $G_\mu$ a large part of
these universal electroweak corrections is absorbed. To describe the
emission of the hard (real) photon, however, we use the
$\alpha(0)-$scheme with $\alpha \equiv \alpha(0) =
1/137$. Consequently, the prediction for the $t\bar{t}\gamma$ cross
section is decreased by more than $3\%$.  Based on the fact that
relative NLO EW corrections to the on-shell $t\bar{t}\gamma$
production at the $13$ TeV LHC are negative and of the order of $2\%$
\cite{Duan:2016qlc} we believe that this is a more consistent approach
compared to employing $\alpha_{G_\mu}$.  In the first
step we use kinematic-independent scales $\mu_R=\mu_F=\mu_0$ with the
central value $\mu_0=m_t/2$ rather than $\mu_0=m_t$. Even though the
mass of the heaviest particle appearing in the process seems to be a
more natural option, this scale choice is motivated by the fact that
$t\bar{t}$ production at the LHC is dominated by $t$-channel gluon
fusion, which favours smaller values of the scale. Additionally, the
contributions beyond NLO that include the resummation of
next-to-leading logarithmic soft gluon effects (NLO+NLL) are smaller
for $\mu_0=m_t/2$ than for $\mu_0=m_t$, as we have explicitly checked
with the help of the \textsc{Top++} program
\cite{Czakon:2011xx}. Taking into account that photon emission is not
a QCD effect this picture should not change for $pp\to t\bar{t}
\gamma$ production. With the goal of stabilising shapes in the high
$p_T$ regions, that are relevant for the new physics searches, we have
explored a dynamical choice for $\mu_R$ and
$\mu_F$. Kinematic-dependent scales should help to achieve flatter
differential ${\cal K}$-factors, thus, to describe more appropriately
regions of the phase-space far away from the $t\bar{t}$ threshold. For
the process at hand, we explored several possibilities and decided in
the end to consider the following dynamical scale
$\mu_R=\mu_F=\mu_0=H_T/4$ where $H_T$ is the total transverse momentum
of the system, which we have defined as
\begin{equation}
H_T= p_{T,\, e^+}+p_{T,\, \mu^-} +p_{T,\, b_1} + p_{T,\, b_2}  +
p_{T}^{miss} + p_{T,\, \gamma} \,,
\end{equation}
where $b_1$ and $b_2$ are bottom-jets (not bottom quarks) and
$p_{T}^{miss}$ is the total missing transverse momentum from escaping
neutrinos. The theoretical uncertainty is estimated with independent
scale variation $\mu_R\ne \mu_F$, subject to the additional restriction
$0.5 < \mu_R/\mu_F < 2$.  In practise such a  restriction amounts to
consider the following pairs
\begin{equation}
\left(\frac{\mu_R}{\mu_0},\frac{\mu_F}{\mu_0}\right) = \left\{
(2,1),(0.5,1),(1,2),(1,1),(1,0.5),(2,2),(0.5,0.5) 
\right\}\,.
\end{equation}
Consequently, the minimum and maximum of the resulting cross section
is chosen. Let us mention here that while calculating the scale
dependence for the NLO cross section we keep $\Gamma_t^{\rm NLO}$
fixed independently of the scale choice. For two scales $\mu =
\mu_0/2$ and $\mu= 2\mu_0$ with $\mu_0 = m_t/2$ the change in the
value of $\Gamma_t^{\rm NLO}$ is smaller than $\pm 1.2\%$. The error
introduced by this treatment is, however, of higher orders. We have
checked that for the simpler case, i.e. for the $pp \to e^+\nu_e \mu^-
\bar{\nu}_\mu b\bar{b} +X $ process, the variation in $\Gamma_t^{\rm
NLO}$ has introduced deviations in the total cross section up to $\pm
1.5\%$ only \cite{Bevilacqua:2010qb}. Let us further note
that in Ref. \cite{Denner:2012yc} a similar procedure has been
discussed. In this paper the mismatch between the scale used in
partial and total top quark decay widths has been compensated by the
so-called partial width correction. The latter has been studied for
total cross section for $pp/p\bar{p} \to e^+ \nu_e \mu^- \bar{\nu}
b\bar{b}+X$ within cuts for Tevatron and LHC at different center of
mass system energies both with fixed and dynamical scales. In the case
of the LHC at $\sqrt{s}=14$ TeV, for example, these partial width
corrections amounted to $1\%-3\%$ depending on the scale choice
employed. At NLO (LO) in QCD we employ CT14nlo (CT14llo)
\cite{Dulat:2015mca} PDFs and describe the running of the strong
coupling constant $\alpha_s$ with two-loop (one-loop) accuracy. The
calculations are performed in the so-called 5 flavour scheme, however,
the $b$-initiated contributions are not taken into account due to
their numerical insignificance. To be more precise already at LO they
are below $0.1\%$. All final-state partons with pseudorapidity $|\eta|
<5$ are recombined into infrared-safe jets via the {\it anti}$-k_T$
jet algorithm \cite{Cacciari:2008gp}. The cone size and jet resolution
parameter $R$ is set to $R=0.4$. We require exactly two $b$-jets, one
photon, two charged leptons and missing transverse momentum,
$p_T^{miss}$.  The hard photon is defined with $p_{T,\,\gamma}>25
~{\rm GeV}$ and $|y_\gamma|<2.5$. To avoid QED collinear singularities
in photon emission, caused by $q \to q \gamma$ splittings, a
separation between quark and photon is required. Since distinguishing
between quark and gluon jets is impossible on the experimental side,
at the same time a separation between photons and gluons is induced as
well. As a consequence, at a given photon $p_T$ an angular restriction
on the soft gluon emission phase-space is introduced. Thus, soft
divergences in the real emission part are different from those in the
virtual correction impairing the cancelation of infrared divergences.
To ensure soft and collinear safety we use a modified cone approach as
described in Ref.~\cite{Frixione:1998jh}, which implements a (smooth)
isolation condition treating quarks and gluons the same way. With the
isolation cone of $R_{\gamma j}=0.4$ for each parton $i$ we evaluate
$\Delta R_{\gamma i}$ between this parton and the photon. We reject
the event unless the following condition is fulfilled
%
\begin{equation}
\sum_{i} E_{T,\,i}  \, \Theta(R - R_{\gamma i})  \le E_{T,\,\gamma} \left(
\frac{1-\cos(R)}{1-\cos(R_{\gamma j})}
\right)
\,,
\label{frixione}
\end{equation}
%
for all $R\le R_{\gamma j}$, where $E_{T,\,i}$ is the transverse
energy of the parton $i$ and $E_{T,\,\gamma}$ is the transverse energy
of the photon. Jets reconstructed inside the cone size $R_{\gamma j}$
are not subject to additional selection criteria.  We apply the
following inclusive cuts to simulate detector response
%
\begin{equation}
\begin{array}{lll}
 p_{T,\,\ell}>30 ~{\rm GeV}\,,    &\quad \quad \quad \quad\quad|y_\ell|<2.5\,,&
\quad \quad \quad \quad \quad
\Delta R_{\ell
 \ell} > 0.4\,,\\[0.2cm]
p_{T,\,b}>40 ~{\rm GeV}\,,  &\quad \quad\quad\quad\quad |y_b|<2.5 \,, 
 &\quad \quad\quad \quad \quad
\Delta R_{bb}>0.4\,,
  \\[0.2cm]
p^{miss}_{T} >20 ~{\rm GeV} \,,   &\quad \quad\quad\quad \quad\Delta R_{\ell
\gamma}>0.4\,,  &\quad \quad \quad
\quad \quad
\Delta R_{\ell b} > 0.4\,, \\
\end{array}
\end{equation}
%
where $\ell$ stands for $\mu^-,e^+$.  We set no restriction on the
kinematics of the extra jet. 

%
\section{Results for the LHC Run II energy of 13 TeV 
for the fixed   scale choice}
\label{sec:4}
%

%
\subsection{Integrated cross section and its scale 
dependence} 
%

%
\begin{figure}[t!]
\begin{center}
\includegraphics[width=0.49\textwidth]{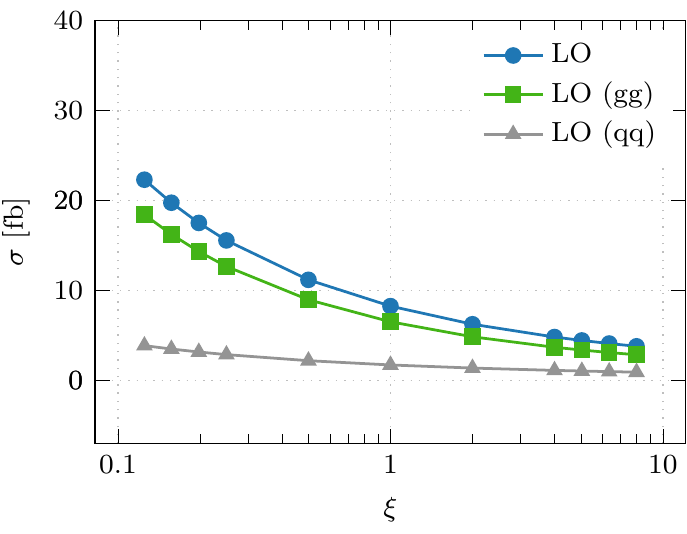}
\includegraphics[width=0.49\textwidth]{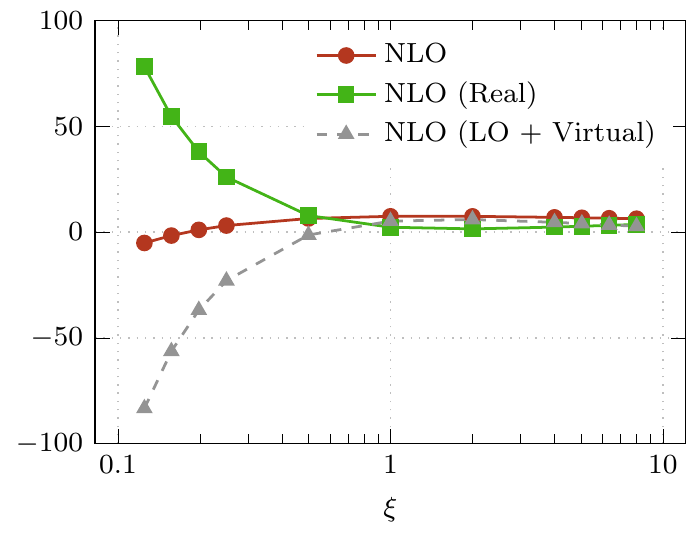}
\includegraphics[width=0.49\textwidth]{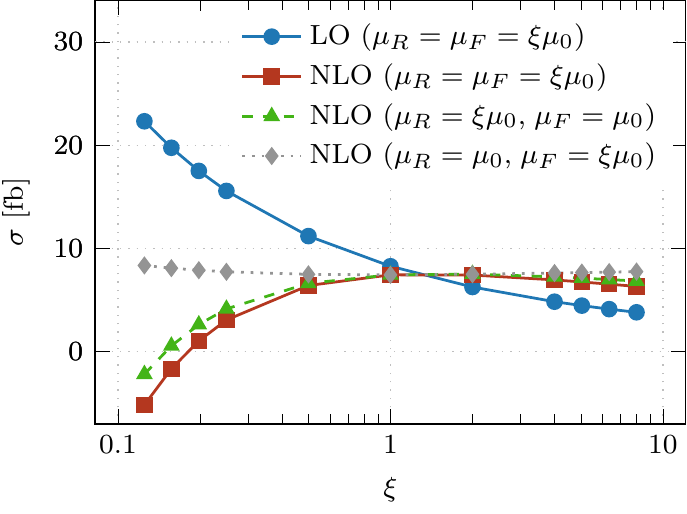}
\end{center}
\vspace{-0.2cm}
\caption{\it Scale dependence of the LO cross section with the
individual contributions of the partonic subprocesses (top-left) 
together with the scale dependence of the NLO cross section 
decomposed into the contribution of the virtual corrections plus LO and real
radiation (top-right). Also shown is the scale dependence of the LO
and NLO integrated cross section obtained by varying $\mu_R$ and
$\mu_F$ simultaneously, as well as NLO scale dependence derived by
varying $\mu_R$ $(\mu_F)$ while  keeping $\mu_F$ $(\mu_R)$ fixed
(bottom). All results are obtained  for $\mu_F=\mu_R=\mu_0$ with
$\mu_0=m_t/2$. The LO and the NLO CT14 PDF sets are employed.}
\label{fig:scale_dependence_mt}
\end{figure}

With the input parameters and cuts specified above, we arrive at
the following  predictions for $\mu_R=\mu_F=\mu_0=m_t/2$
%
\begin{equation}
\begin{split}
\sigma^{\rm LO}_{pp\to e^+\nu_e\mu^-\bar{\nu}_\mu b\bar{b}\gamma}
({\rm CT14}, \mu_0=m_t/2) &=
8.27{~}^{+2.92 \, (35\%)}_{-2.01 \,(24\%)}  ~ \text{fb}\,, \\[0.2cm]
\sigma^{\rm NLO}_{pp\to e^+\nu_e\mu^-\bar{\nu}_\mu b\bar{b}\gamma} 
({\rm CT14}, \mu_0=m_t/2)&=
7.44{~}^{+0.07\, (\,\,\,1\%)}_{-1.03\, (14\%)} ~\text{fb}\,.
\end{split}
\end{equation}
%
At the central scale $\mu_0=m_t/2$, the $gg$ channel dominates the
total LO $pp$ cross section by $79\%$, followed by the
$q\bar{q}+\bar{q}q$ channel with $21\%$. Photons are, therefore,
predominantly radiated off the top quark and top quark decay
products. The full $pp$ cross section receives negative and moderate
NLO corrections of $10\%$. The theoretical uncertainties resulting
from scale variations, where $\mu_R$ and $\mu_F$ have been varied
independently, and taken in a very conservative way as a
maximum of the lower and upper bounds are $35\%$ at LO and $14\%$ at
NLO. Thus, a reduction of the theoretical error by a factor of $2.5$
is observed. Should we instead vary $\mu_R$ and $\mu_F$
simultaneously, up and down by a factor of $2$ around $\mu_0$, the
uncertainties would remain unchanged. This is due to the fact that the
scale variation is driven solely by the changes in $\mu_R$. In the
case of truly asymmetric uncertainties, however, it is always more
appropriate to symmetrise the errors. After symmetrisation the scale
uncertainty at LO is assessed to be instead of the order of
$30\%$. After inclusion of the NLO QCD corrections, the scale
uncertainty is reduced down to $7\%$. The graphical presentation of
the behaviour of LO and NLO cross sections upon varying the scale by a
factor $\xi \in \left\{0.125,\dots,8\right\}$ is shown in
Fig.~\ref{fig:scale_dependence_mt}.  At LO the individual
contributions of the partonic subprocesses are additionally presented.
The final scale dependence of the NLO cross section as emerged
out of the two contributions (the virtual plus the LO part and the
real emission part) is also
depicted in Fig. ~\ref{fig:scale_dependence_mt}.  Of course, the
separation is entirely unphysical, but well defined once we state that
we use the 't Hooft-Veltman version of the dimensional regularisation,
with the integrals as defined in the \textsc{OneLOop} library.

Next, we have checked the dependence on the parameters introduced 
in the photon isolation procedure. Specifically,  the general photon
isolation formula is given  by 
%
\begin{equation}
\sum_{i} E_{T,\,i}  \, \Theta(R - R_{\gamma i})  \le
\epsilon_\gamma\,  \,  E_{T,\,\gamma} \left(
\frac{1-\cos(R)}{1-\cos(R_{\gamma j})}
\right)^n
\,,
\end{equation}
%
with two additional parameters $\epsilon_\gamma$ and $n$.  The default
choice, which should guarantee moderate corrections, is
$\epsilon_\gamma=1$ and $n=1$, see
Ref.~\cite{Frixione:1998jh}. Nevertheless, both $\epsilon_\gamma$ and 
$n$ a priori can have arbitrary values. We have recalculated the
subtracted real emission part of the  NLO results with a different
choice, namely $\epsilon_\gamma=1/2$ and $n=1/2$. Within the
integration errors our new results have agreed with the old ones. Thus, 
NLO QCD results for the $pp\to e^+\nu_e \mu^-\bar{\nu}_\mu b\bar{b}
\gamma +X$ production process are not sensitive to moderate changes in
values of  $\epsilon_\gamma$ and  $n$. 

In the subsequent step, the size of the top quark non-factorisable
corrections has been estimated for the total cross section. To achieve
this the full result has been compared with the result in the NWA.
The latter has been obtained by rescaling the coupling of the top
quark to the $W$ boson and the $b$ quark by several large factors, 
as described in Ref. \cite{Bevilacqua:2010qb}, to mimic
the limit $\Gamma_t \to 0$ when the scattering cross section
factorizes into on-shell production and decay. The top quark
non-factorisable corrections for the $pp\to e^+\nu_e
\mu^-\bar{\nu}_\mu b\bar{b} \gamma +X$ production process amount to
$1.5\%$ $(2.5\%)$ for LO (NLO). They are consistent with the expected
uncertainty of the NWA, which is of the order of ${\cal
O}(\Gamma_t/m_t)$.

Coming back to the theoretical uncertainties, we note that, another
source of theoretical uncertainties comes from the PDF
parametrisation. To that end, we have recomputed NLO QCD corrections
to the $pp\to e^+ \nu_e \mu^-\bar{\nu}_\mu b\bar{b} \gamma +X$
production process with different PDF sets. Following recommendations
of PDF4LHC for the usage of PDFs suitable for applications at the LHC
Run II \cite{Butterworth:2015oua} we employ additionally to the CT14
PDF set the MMHT14 PDF set \cite{Harland-Lang:2014zoa} and NNPDF3.0
\cite{Ball:2014uwa}. Let us say here, that differences coming from
NLO results for various PDF sets are comparable (usually even higher)
to the individual estimates of PDF systematics.  We have checked that
this is the case for the similar process, namely for $pp \to e^+
\nu_e\mu^- \bar{\nu}_\mu b\bar{b} j +X$ production
\cite{Bevilacqua:2016jfk}. In this paper, we take the PDF uncertainties
to be the difference between our default PDF set (CT14) and the other
two PDF sets considered (MMHT14 and NNPDF3.0).
Our findings for  
MMHT14 and NNPDF3.0 PDF  can be summarised as follows
%
\begin{equation}
\begin{split}
\sigma^{\rm NLO}_{pp\to e^+\nu_e\mu^-\bar{\nu}_\mu b\bar{b}\gamma}
({\rm MMHT14}, \mu_0=m_t/2) &= 7.49 ~{\rm fb}\,,
\\[0.2cm]
\sigma^{\rm NLO}_{pp\to e^+\nu_e\mu^-\bar{\nu}_\mu b\bar{b}\gamma} 
({\rm NNPDF3.0}, \mu_0=m_t/2)&=  7.72 ~{\rm fb}\,.
\end{split}
\end{equation}
%
The PDF uncertainties for the process under scrutiny are, therefore, given by
$+0.05 ~{\rm fb} ~(1\%)$ for the {\rm MMHT14} PDF set and $+0.28
~{\rm fb} ~(4\%)$ for {\rm NNPDF3.0}. Our result for the integrated
cross section at NLO in QCD with the CT14 PDF set and for
$\mu_0=m_t/2$ is given by
\begin{equation}
\sigma^{\rm NLO}_{pp\to e^+\nu_e\mu^-\bar{\nu}_\mu b\bar{b}\gamma} 
({\rm CT14}, \mu_0=m_t/2)=
7.44{~}^{+0.07\, (\,\,\,1\%)}_{-1.03\, (14\%)} \, [{\rm scales}] \,
{}^{+0.05 \, (1\%)}_{+0.28\,  (4\%)} \, [{\rm PDF}]~ \text{fb}\,.
\end{equation}
Taken in a very conservative way, the PDF uncertainties are of the
order of $4\%$ (to be compared to the theoretical uncertainties of
$14\%$ from the scale dependance). After symmetrisation they are
reduced down to $2\%$ (to be compared to $7\%$). Overall, the PDF
uncertainties are well below the theoretical uncertainties due to the
scale dependence. The latter remain the dominant source of the
theoretical systematics.

%
\subsection{Differential cross sections}
%

\begin{figure}[t!]
\begin{center}
\includegraphics[width=0.49\textwidth]{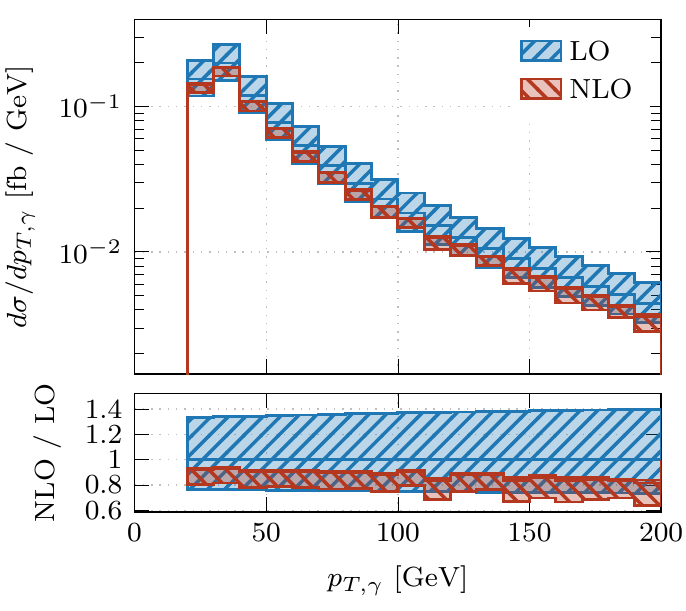}
\includegraphics[width=0.49\textwidth]{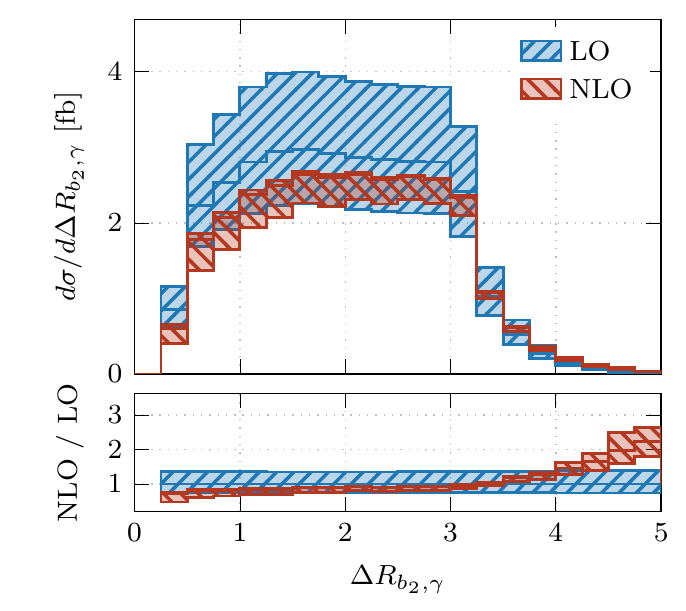}
\end{center}
\vspace{-0.2cm}
\caption{\it
Differential distributions as a function of the transverse momentum of
the hard photon, $p_{T,\gamma}$ and the rapdity-azimuthal angle
separation between the photon and the softer $b$-jet, $\Delta
R_{b_2,\gamma}$, for $\mu_F=\mu_R=\mu_0=m_t/2$. The LO and the NLO
CT14 PDF sets are employed. The upper panels show absolute LO and NLO
predictions together with corresponding uncertainty bands. The lower
panels display the differential ${\cal K}$-factor together with the
uncertainty band and the relative scale uncertainties of the LO cross
section.}
\label{fig:distribution1_mt}
\end{figure}
%
\begin{figure}[t!]
\begin{center}
\includegraphics[width=0.49\textwidth]{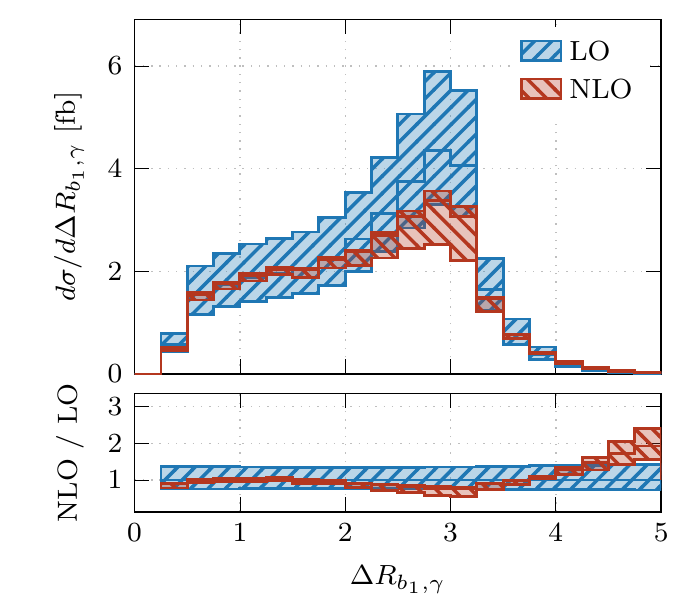}
\includegraphics[width=0.49\textwidth]{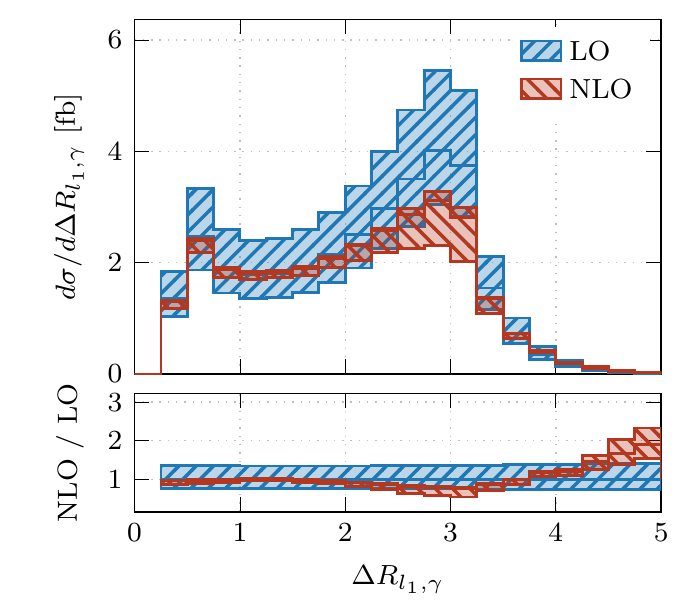}
\includegraphics[width=0.49\textwidth]{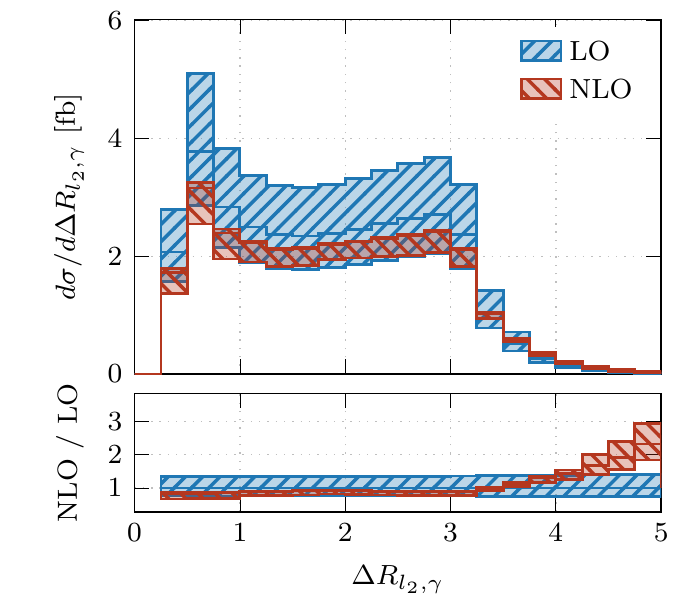}
\end{center}
\vspace{-0.2cm}
\caption{\it
Differential distributions as a function of the separation in the
rapidity-azimuthal angle plane between the hard photon and the hardest
b-jet, $\Delta R_{b_1, \gamma}$, as well as the separation between the
hard photon and the hardest and softer lepton, $\Delta R_{\ell_1,
\gamma}$ and $\Delta R_{\ell_2, \gamma}$ for
$\mu_F=\mu_R=\mu_0=m_t/2$. The LO and the NLO CT14 PDF sets are
employed. The upper panels show absolute LO and NLO predictions
together with corresponding uncertainty bands. The lower panels
display the differential ${\cal K}$-factor together with the
uncertainty band and the relative scale uncertainties of the LO cross
section.}
\label{fig:distribution2_mt}
\end{figure}
%
\begin{figure}[t!]
\begin{center}
\includegraphics[width=0.49\textwidth]{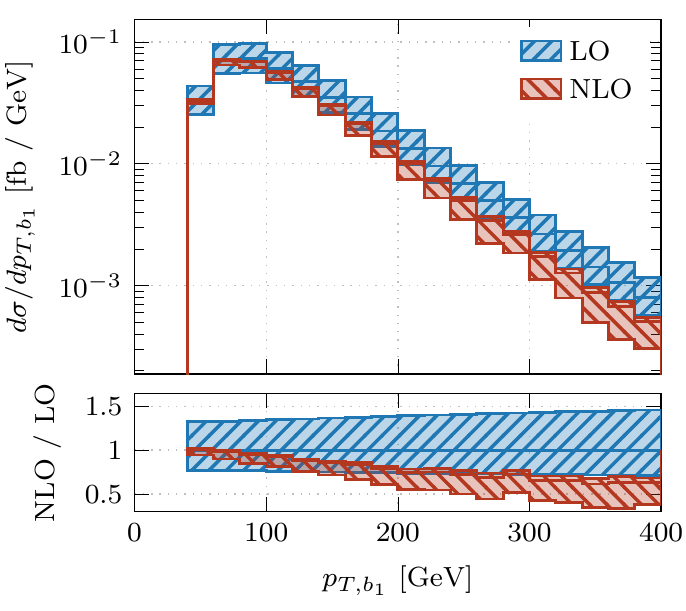}
\includegraphics[width=0.49\textwidth]{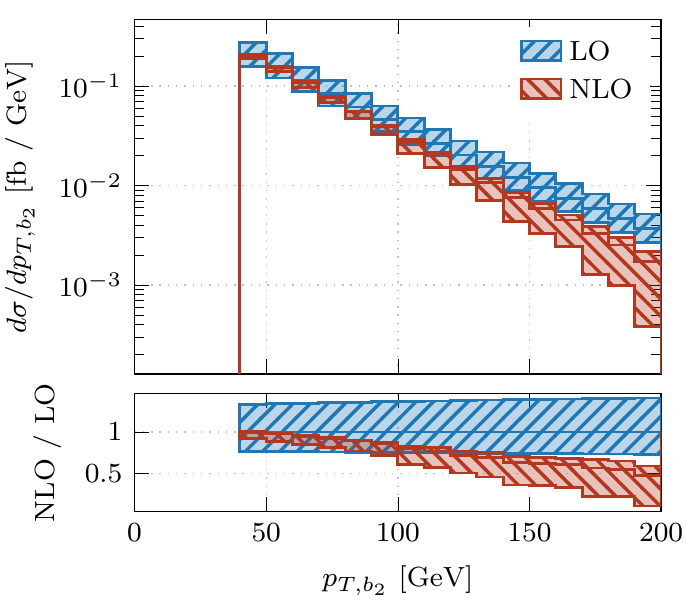}
\includegraphics[width=0.49\textwidth]{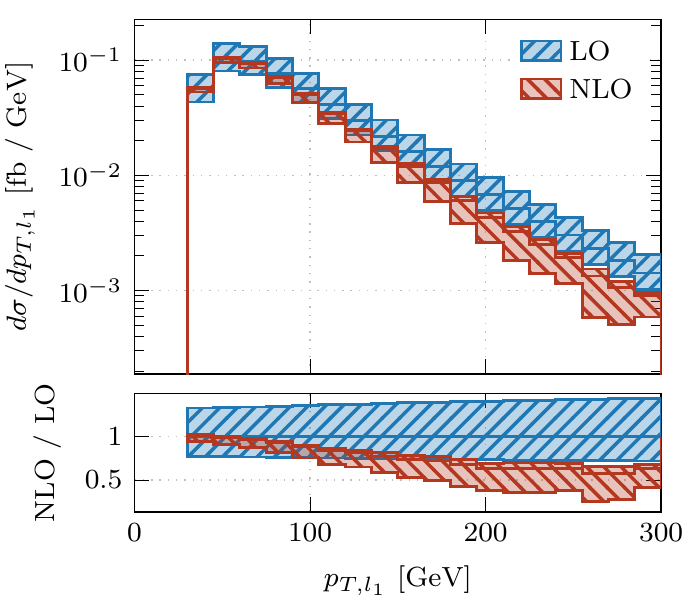}
\includegraphics[width=0.49\textwidth]{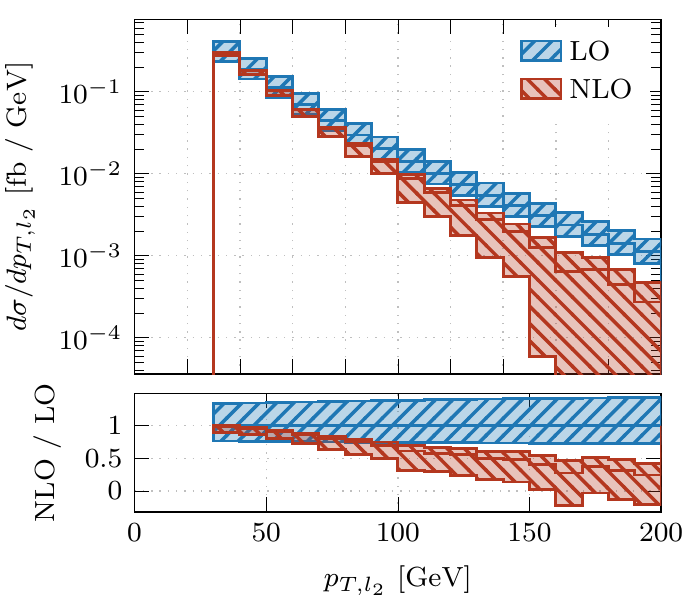}
\end{center}
\vspace{-0.2cm}
\caption{\it
Differential distributions as a function of $p_T$ of the hardest and
the softer $b$-jet as well as the hardest and the softer lepton for
$\mu_F=\mu_R=\mu_0=m_t/2$. The LO and the NLO CT14 PDF sets are
employed.  The upper panels show absolute LO and NLO predictions
together with corresponding uncertainty bands. The lower panels
display the differential ${\cal K}$-factor together with the
uncertainty band and the relative scale uncertainties of the LO cross
section.}
\label{fig:distribution3_mt}
\end{figure}

While the size of higher order corrections to the total cross section
is certainly interesting, it is crucial to study the corrections to
differential distributions. In Fig.~\ref{fig:distribution1_mt} we present
representative differential distributions, that are relevant for BSM
searches \cite{Baur:2001si,Baur:2004uw}. We display $p_{T}$ of the
hard photon and $\Delta R_{b_2, \gamma}$ between the hard photon and
the softer $b$-jet. The upper panels show the distributions themselves
and their scale dependence. The lower panels reveal the differential
${\cal K}$-factor with its error and the relative scale uncertainties
of the LO cross section.  To be more precise we plot ${\cal K}^{\rm
NLO}(\mu)=(d\sigma^{\rm NLO}(\mu)/dX)/(d\sigma^{\rm LO}(\mu_0)/dX)$
and ${\cal K}^{\rm LO}(\mu)=(d\sigma^{\rm LO}(\mu)/dX)/(d\sigma^{\rm
LO}(\mu_0)/dX)$ where $\mu_0=m_t/2$ is the central value of the scale
and $X$ denotes the observable that is scrutinised.  Higher order
corrections have strongly altered the shape of $\Delta R_{b_2,
\gamma}$ where corrections range from $-29\%$ to $+122\%$, causing
distortions of up to $150\%$. Similar effects have been noticed for
other observables, most notably for other angular observables 
shown in Fig.~\ref{fig:distribution2_mt}. Among
others the most affected by higher order corrections are the
separation in the rapidity-azimuthal angle plane between the hard
photon and the hardest $b$-jet, $\Delta R_{b_1, \gamma}$ (NLO
corrections from $-24\%$ to $+93\%$), as well as the separation
between the hard photon and the hardest or softer charged lepton,
$\Delta R_{\ell_1, \gamma}$ (NLO corrections ranging from $-25\%$ to $+91\%$)
and $\Delta R_{\ell_2, \gamma}$ (NLO corrections ranging from $-16\%$ to
$+132\%$).  In each case the large differential ${\cal K}$-factor for $\Delta
R \gtrsim 4 $ is associated with photon emission from
initial state quark from the $qg+gq$ partonic subprocess, where $q$
stands for quark and antiquark. Such a contribution appears only
starting at NLO and adds significantly at large $\Delta R$. Let
us mention here, that the $qg+gq$ channel contribution to $\sigma^{\rm
NLO}_{pp\to e^+\nu_e \mu^-\bar{\nu}_\mu b\bar{b}\gamma}$ is estimated
at the level of $29\%$.  Moreover, due to the leading order like
nature of the contribution also the scale dependence in this region is
enlarged. Let us additionally note here, that emission of the
photon from the charged lepton leads to collinear enhancement at small
values of the separation between the photon and the softer charged
lepton, $\Delta R_{\ell_2,\gamma}$ as can be clearly observed in
Fig.~\ref{fig:distribution2_mt}. Moreover, in the case of the
separation between the photon and the softer $b$-jet,
$R_{b_2,\gamma}$, depicted in Fig.~\ref{fig:distribution1_mt},
 events are produced over a wide range of $\Delta R_{b_2,\gamma}$
 values rather than in the back-to-back configuration. This confirms
 the findings of Ref.~\cite{Melnikov:2011ta} that photon radiation off
 top quark decay products yields a significant contribution to the
 cross-section.

 In case of $p_{T,\gamma}$ the differential ${\cal K}$-factor is rather
constant  only in the plotted range from $-8\%$ to $-18\%$.
Thus, $p_{T}$ of the photon is more stable against higher order
corrections and hence better suited for BSM searches. Nevertheless,
both observables require higher order calculations to be properly
described. In view of ongoing indirect searches for BSM physics, where
the emphasis is looking for small deviations from the most accurate
SM predictions, such state of the art results are indispensable.

Finally, in Fig.~\ref{fig:distribution3_mt} we
present dimensionful observables. Specifically, we display
the transverse momentum distributions of the hardest and the softer
$b$-jet and charged lepton. In all cases negative and large higher
order QCD corrections can be detected. In the high $p_T$ regions they
amount to $-38\%$, $-53\%$, $-43\%$ and $-76\%$ for $p_{T,\, b_1}$,
$p_{T,\, b_2}$, $p_{T,\, \ell_1}$ and $p_{T,\, \ell_2}$
respectively. Moreover, the NLO error bands do not fit within the LO
ones as one would expect from a well-behaved perturbative expansion.
Thus, the fixed scale choice does not ensure a stable shape when going
from LO to NLO for these observables. Through the implementation of a
dynamical scale, the large discrepancies between the shapes of these
distributions at NLO and LO should disappear. Thus, in the next step
we shall examine NLO results for $\mu_R=\mu_F=\mu_0=H_T/4$ with the
goal of stabilising differential ${\cal K}$-factor, i.e. decreasing 
NLO QCD corrections in the tails, for $p_{T,\, b_1}$, $p_{T,\, b_2}$, 
$p_{T,\,\ell_1}$ and $p_{T,\, \ell_2}$ while keeping the behaviour 
of ${\cal K}$ almost unchanged for $p_{T,\, \gamma}$.

%
\subsection{Theoretical uncertainties for differential 
cross sections}
%

\begin{figure}[t!]
\begin{center}
\includegraphics[width=0.49\textwidth]{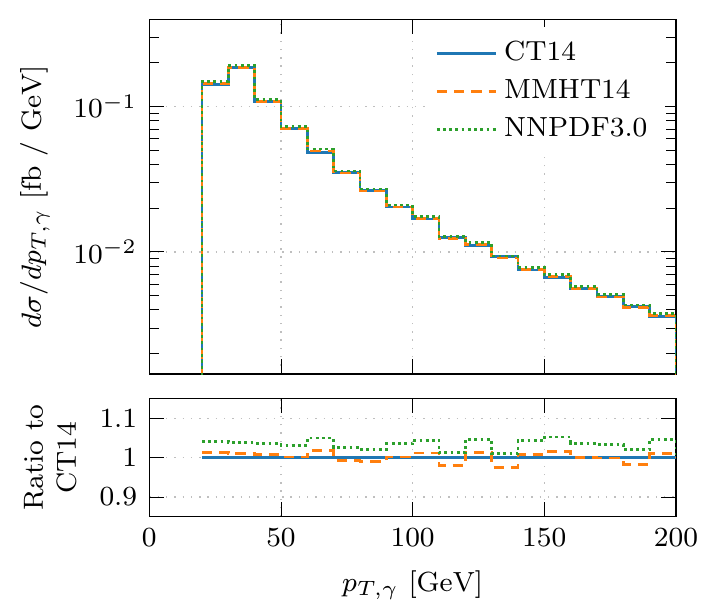}
\includegraphics[width=0.49\textwidth]{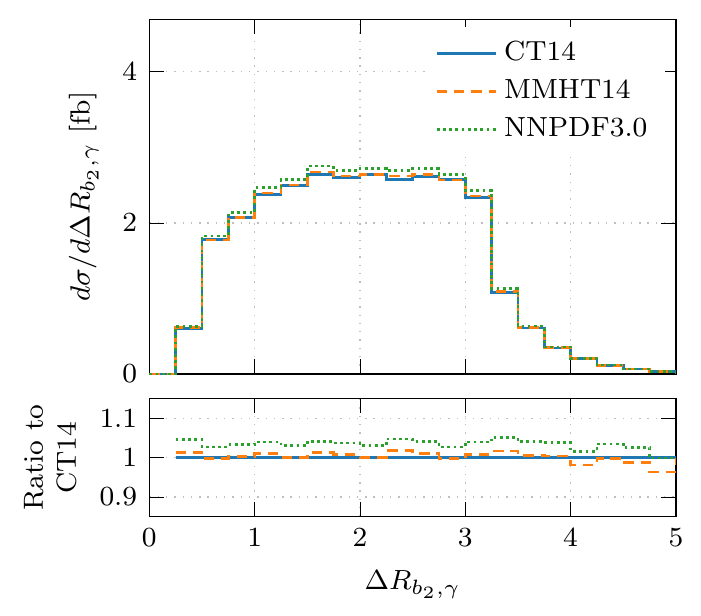}
\end{center}
\vspace{-0.2cm}
\caption{\it  
 NLO differential distributions as a function of the
transverse momentum of the hard photon and the separation between the
photon and the softer b-jet, $\Delta R_{b_2,\gamma}$.  Results are
shown for $\mu_F = \mu_R = \mu_0 = m_t/2$ and for three different PDF
sets. Lower panels display the ratio of the
MMHT14 (NNPDF3.0) PDF set to CT14.}
\label{fig:pdfs_mt}
\end{figure}

At this point we would like to estimate theoretical uncertainties
inherent in our LO and NLO differential cross sections as obtained
with $\mu_0=m_t/2$ and the CT14 PDF set. The scale uncertainties are
again estimated conservatively by scanning bin by bin values of the
lower and upper bounds and by choosing the maximal number. To get a
general idea about the size of theoretical errors we quote here only
this maximal value. In this way, for the transverse momentum
distribution of the hard photon we have obtained theoretical errors up
to $\pm 40\%$ at LO and up to $\pm 22\%$ at NLO. For dimensionful
observables these maximal values come from the high $p_T$ regions. A
similar pattern can be seen for the angular separation between the
hard photon and $b$-jet or the charged lepton. Specifically, for
$\Delta R_{b_2,\gamma}$ we have $\pm 40\%$ at LO to be compared with
$\pm 33\%$ at NLO and for $\Delta R_{b_1,\gamma}$ is $\pm 42\%$ at LO
versus $\pm 28\%$ at NLO. In case of the charged lepton the situation is
very similar as we have estimated the theoretical error at the level
of $\pm 42\%$ ($\pm 40\%$) at LO and $\pm 28\%$ ($\pm 27\%$) at NLO
for $\Delta R_{\ell_1,\gamma}$ ($\Delta R_{\ell_2,\gamma}$). Thus, in
all above mentioned cases a reduction by a factor of $1.5-2$ is
achieved by increasing the order in perturbative expansion. However, in
case of transverse momentum distributions of $b$-jets and leptons the
picture has changed and there is a large residual scale dependence in
these observables even at NLO. Actually, for the $p_{T,b_1}$
distribution the theoretical error is at the same level independently of
the perturbative order and amounts to $\pm 46\%$. For the $p_{T,\ell_1}$
and $p_{T,b_2}$ at NLO the theoretical error is larger than at LO,
respectively $\pm 56\%$ and $\pm 78\%$. Finally, for the last plotted
observable, i.e.  $p_{T,\ell_2}$ huge uncertainties of the order of
$\pm 186\%$ can be seen. This clearly tell us that $\mu_0=m_t/2$ is
not equipped to properly describe tails of $p_T$ distributions even at
NLO.  Many of these features can be improved by performing NLO
computation with the kinematic-dependent choices of the scales.

Lastly, we have examined PDF uncertainties for the differential cross
sections with the fixed scale choice. For all observables that we have
studied PDF uncertainties are negligible in comparison to the
theoretical uncertainties from the scale dependence. As an example we
show in Fig.~\ref{fig:pdfs_mt} NLO differential 
distributions as a function of the transverse momentum of the hard
photon and the azimuthal angle-rapidity distance between the hard
photon and the softest $b$-jet. The upper panels present the NLO
predictions for three different PDF sets at the central scale value
$\mu_R=\mu_F=\mu_0=m_t/2$.  In addition to the CT14 PDF set, we employ
the MMHT14 and NNPDF3.0 PDF sets.  The lower panels of
Fig.~\ref{fig:pdfs_mt} give the ratio of the MMHT14 (NNPDF3.0) PDF set
to CT14.

To summarise this part, for $pp \to e^+ \nu_e \mu^- \bar{\nu}_\mu
b\bar{b}\gamma +X$ production at the LHC Run II with $\sqrt{s}=13$ TeV
with our selection of cuts and input parameters, the PDF uncertainties
are insignificant both at the level of total and differential cross
sections once contrasted with theoretical errors from the scale
dependence.  Let us note at this point, however, that
additional theoretical effects should be examined for the process at
hand.  These include among others NLO electroweak effects, the size
of which has to be estimated and compared to the size of NLO QCD effects.
Moreover, dedicated analyses of complete NLO QCD off-shell effects of
the top quark at the differential level have to be carried out. We
leave both aspects for  future studies. Thus, from now on we shall
concentrate only on theoretical uncertainties from the scale
dependence.

%
\section{Results for the LHC Run II energy of 13 TeV 
for  the dynamical   scale choice}
\label{sec:5}
%

%
\subsection{Integrated cross section and its scale 
dependence}
%

\begin{figure}[t!]
\begin{center}
\includegraphics[width=0.49\textwidth]{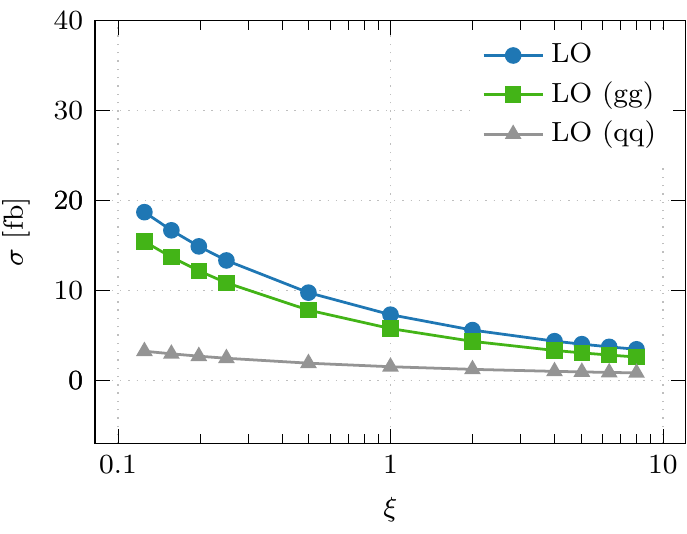}
\includegraphics[width=0.49\textwidth]{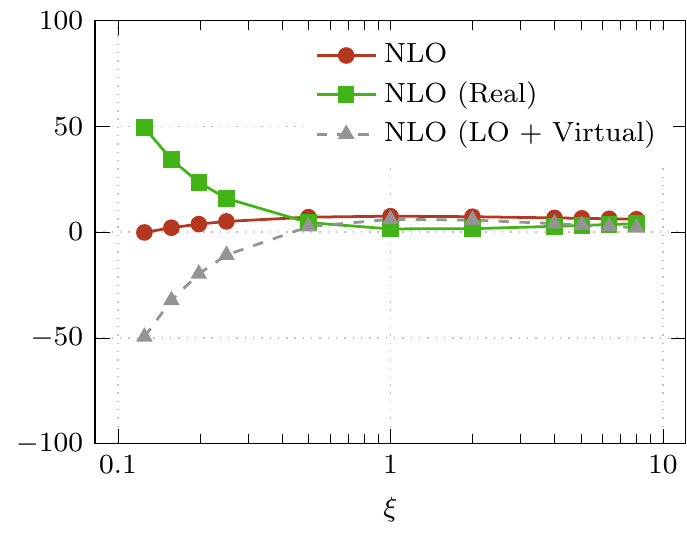}
\includegraphics[width=0.49\textwidth]{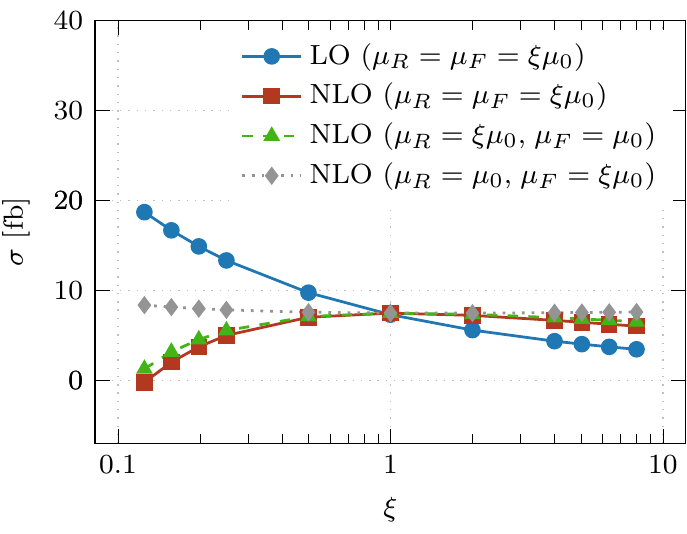}
\end{center}
\vspace{-0.2cm}
\caption{\it 
Scale dependence of the LO cross section with the
individual contributions of the partonic subprocesses together with
scale dependence of the NLO order cross section decomposed into the
contribution of the virtual corrections plus LO and real
radiation. Also shown is scale dependence of the LO and NLO integrated
cross section and the variation of $\mu_R$ $(\mu_F)$ with fixed $\mu_F
$ $(\mu_R)$. All results are obtained for $\mu_F=\mu_R=\mu_0$ with
$\mu_0=H_T/4$. The LO and the NLO CT14 PDF sets are employed.}
\label{fig:scale_dependence_ht}
\end{figure}

For the kinematic-dependent scale $\mu_R=\mu_F=\mu_0=H_T/4$ our
results can be summarised as follows 
%
\begin{equation}
\begin{split}
\sigma^{\rm LO}_{pp\to e^+\nu_e\mu^-\bar{\nu}_\mu b\bar{b}\gamma}
({\rm CT14}, \mu_0=H_T/4) &=
7.32{~}^{+2.44\, (33\%)}_{-1.71 \,(23\%)}  ~ \text{fb}\,, \\[0.2cm]
\sigma^{\rm NLO}_{pp\to e^+\nu_e\mu^-\bar{\nu}_\mu b\bar{b}\gamma} 
({\rm CT14}, \mu_0=H_T/4)&=
7.50{~}^{+0.10\, (1\%)}_{-0.46\, (6\%)} ~\text{fb}\,.
\end{split}
\end{equation}
%
As expected they are in agreement with results provided at LO and at
NLO for $\mu_0=m_t/2$. Precisely, within quoted theoretical errors
they agree at the level of $0.2\sigma$ at LO and $0.05\sigma$ at NLO.
This time, however, the full $pp$ cross section receives positive and
small NLO corrections of $2.5\%$. The theoretical uncertainties
resulting from scale variations are $33\%$ at LO and $6\%$ at NLO. A
reduction of the theoretical error by a factor of $5.5$ is observed
for $\mu_0=H_T/4$. After symmetrisation of theoretical errors the
scale uncertainty at LO is estimated to be instead of the order of
$28\%$ and at NLO is reduced down to $4\%$. Therefore, by going from
LO to NLO we have reduced theoretical error by a factor of $7$. The
graphical display of scale dependence is shown in
Fig.~\ref{fig:scale_dependence_ht}. The new scale choice indeed
captures parts of unknown higher order corrections. After all not only
the size of NLO corrections is diminished but also the theoretical
error is smaller when comparing to the results with the fixed scale
choice.  Scale dependence of the LO cross section with the individual
contributions of the partonic subprocesses and scale dependence of the
NLO order cross section decomposed into the contribution of the
virtual corrections plus LO and the real radiation part are
additionally given in Fig.~\ref{fig:scale_dependence_ht}.  Moreover,
the variation of $\mu_R$ $(\mu_F)$ with fixed $\mu_F $ $(\mu_R)$ is
presented in Fig.~\ref{fig:scale_dependence_ht} as well. Here
qualitatively our findings remain the same as for the fixed scale
choice.

%

%
\subsection{Differential cross sections}
%

\begin{figure}[t!]
\begin{center}
\includegraphics[width=0.49\textwidth]{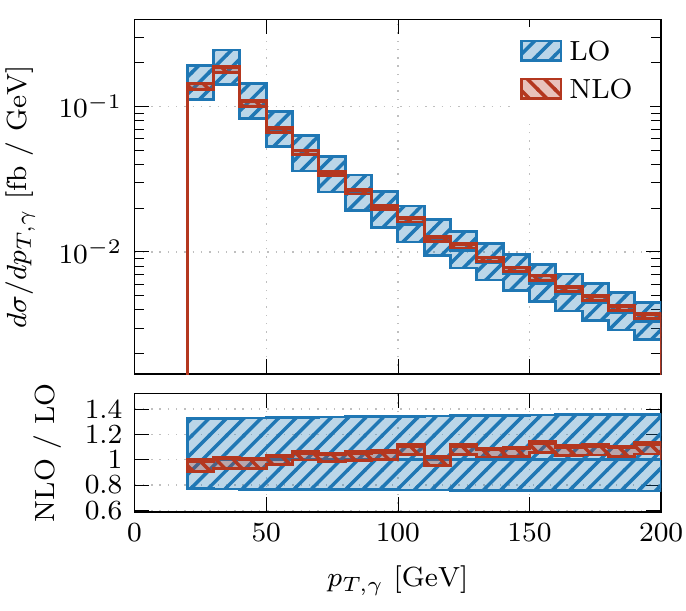}
\includegraphics[width=0.49\textwidth]{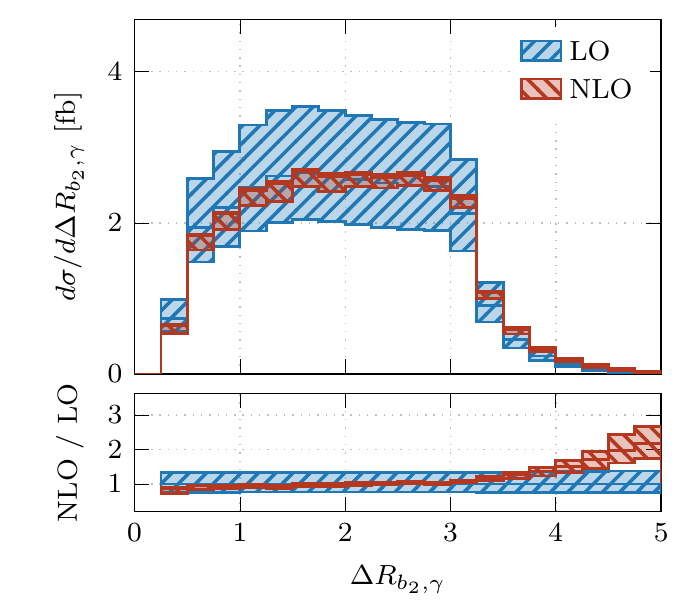}
\end{center}
\vspace{-0.2cm}
\caption{\it
Differential distributions as a function of $p_T$ of the hard photon
and $\Delta R_{b_2\,,\gamma}$ between the photon and the softer
$b$-jet for $\mu_F=\mu_R=\mu_0=H_T/4$. The LO and the NLO CT14 PDF
sets are employed. The upper panels show absolute LO and NLO
predictions together with corresponding uncertainty bands. The lower
panels display the differential ${\cal K}$-factor together with the
uncertainty band and the relative scale uncertainties of the LO cross
section.}
\label{fig:distribution1_ht}
\end{figure}
%
\begin{figure}[t!]
\begin{center}
\includegraphics[width=0.49\textwidth]{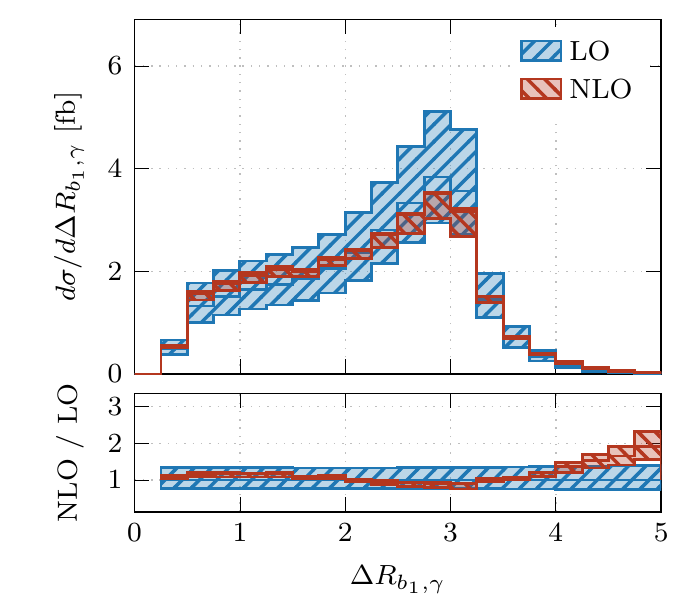}
\includegraphics[width=0.49\textwidth]{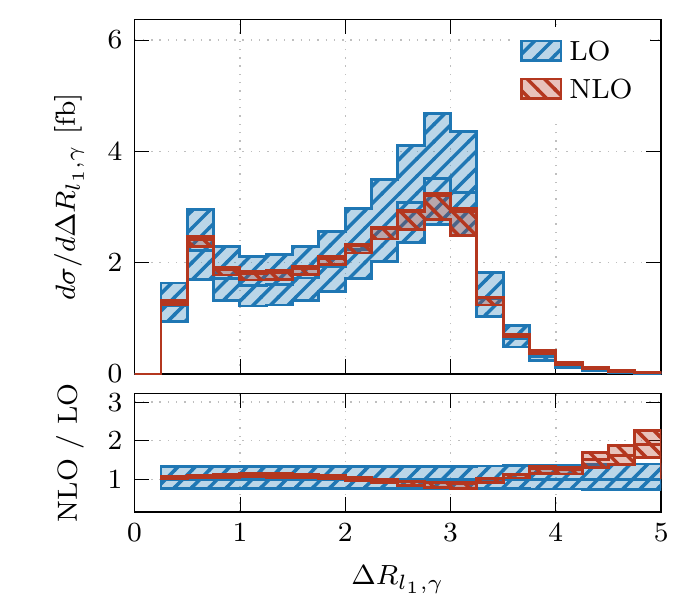}
\includegraphics[width=0.49\textwidth]{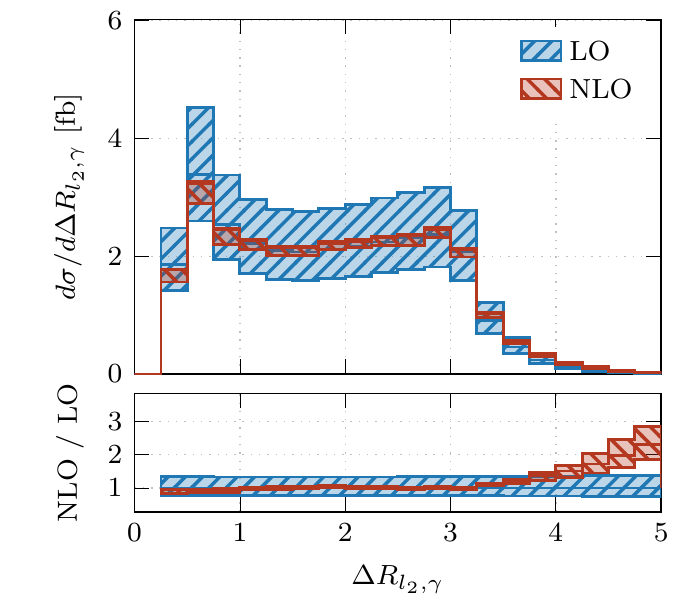}
\end{center}
\vspace{-0.2cm}
\caption{\it
Differential distributions as a function of the separation in the
rapidity-azimuthal angle plane between the hard photon and the hardest
b-jet, $\Delta R_{b_1, \gamma}$, as well as the separation between the
hard photon and the hardest and softer charged lepton, $\Delta
R_{\ell_1, \gamma}$ and $\Delta R_{\ell_2, \gamma}$ for
$\mu_F=\mu_R=\mu_0=H_T/4$.  The LO and the NLO CT14 PDF sets are
employed. The upper panels show absolute LO and NLO predictions
together with corresponding uncertainty bands. The lower panels
display the differential ${\cal K}$-factor together with the
uncertainty band and the relative scale uncertainties of the LO cross
section.}
\label{fig:distribution2_ht}
\end{figure}
%
\begin{figure}[t!]
\begin{center}
\includegraphics[width=0.49\textwidth]{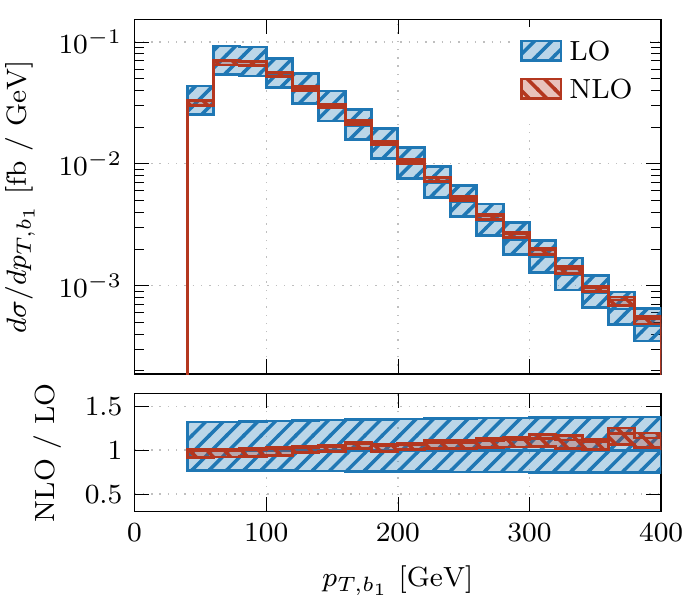}
\includegraphics[width=0.49\textwidth]{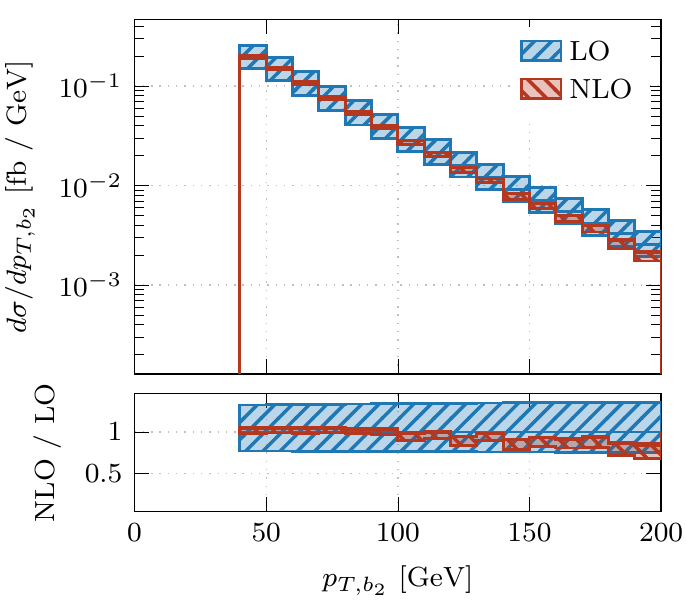}
\includegraphics[width=0.49\textwidth]{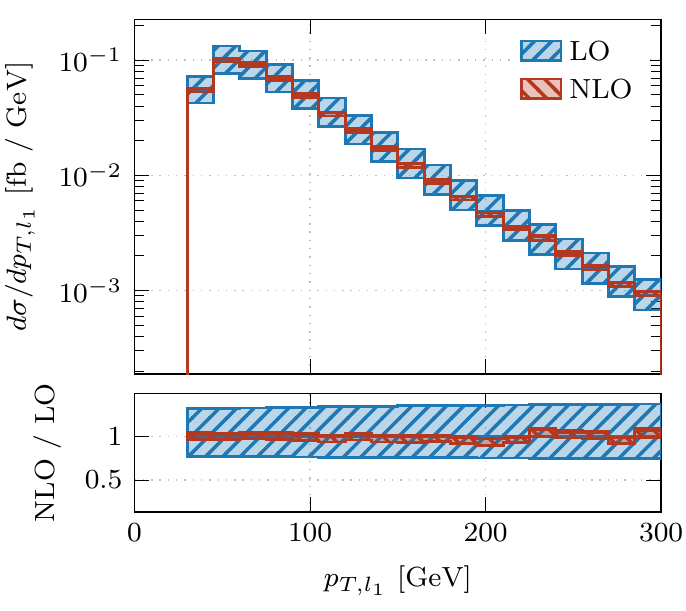}
\includegraphics[width=0.49\textwidth]{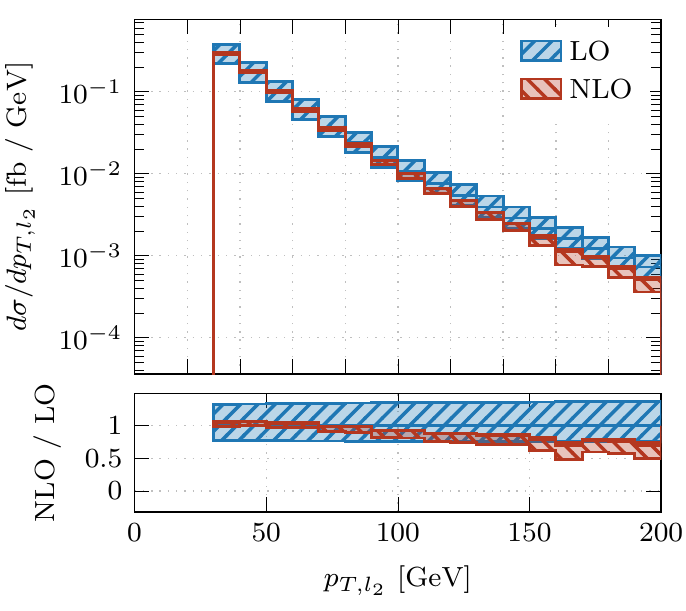}
\end{center}
\vspace{-0.2cm}
\caption{\it
Differential distributions as a function of $p_T$ of the
hardest and the softer lepton as well as the hardest and the softer
$b$-jet for $\mu_F=\mu_R=\mu_0=H_T/4$. The LO and the NLO CT14 PDF
sets are employed.  The upper panels show absolute
LO and NLO predictions together with corresponding uncertainty
bands. The lower panels display the differential ${\cal K}$-factor
together with the uncertainty band and the relative scale
uncertainties of the LO cross section.}
\label{fig:distribution3_ht}
\end{figure}

We turn now our attention to differential cross sections
for $\mu_0=H_T/4$. We have examined the same set of observables as in
the case of $\mu_0=m_t/2$.  Our goal being to find flatter results for
differential ${\cal K}$ factors for dimensionful observables without
introducing major changes in the differential ${\cal K}$ factor of
$p_{T,\,\gamma}$. In the case of $p_{T,\,\gamma}$ already for
$\mu_0=m_t/2$ quite stable (negative) corrections have been
observed. Specifically, shape distortions up to only $10\%$ have been
detected. 

We start with differential distribution for the transverse momentum of
the hard photon that is displayed in
Fig.~\ref{fig:distribution1_ht}. For the dynamical scale choice of
$\mu_0=H_T/4$ positive corrections up to $13\%$ are obtained. We can
also notice that the NLO error band as calculated through the scale
variation is within the LO error band as it should be for a
well behaved observable that is described by the perturbative
expansion in $\alpha_s$. For the dimensionless observable $\Delta
R_{b_2, \gamma}$, that is also shown in
Fig.~\ref{fig:distribution1_ht}, the size of NLO corrections has been
moderately reduced. The higher order corrections range now from
$-12\%$ up to $+116\%$. Thus, the shape distortion up to $128\%$ has
been obtained for this observable, which should be compared with
$150\%$ for the fixed scale choice. Other dimensionless observable are
presented in Fig.~\ref{fig:distribution2_ht}.  For the differential
distribution as a function of the separation in the rapidity-azimuthal
angle plane between the hard photon and the $b$-jet or lepton,
i.e. $\Delta R_{b_1, \gamma}$, $\Delta R_{\ell_1, \gamma}$ and $\Delta
R_{\ell_2, \gamma}$ we have acquired NLO QCD corrections in the
following range $\left\{-11\% , +90\%\right\}$,
$\left\{-10\%,+90\%\right\}$ and $\left\{-5\%,+130\%\right\}$
respectively. In each case shape distortions have been decreased by
about $15\%-20\%$.

Finally, we have reexamined dimensionful observables. Specifically,
transverse momentum distributions of the hardest and the softer lepton
as well as transverse momentum distributions of the hardest and the
softer $b$-jet. They are given in
Fig.~\ref{fig:distribution3_ht}. Higher order corrections in high
$p_T$ regions have been substantially reduced. For the transverse
momentum distribution of the hardest $b$-jet we have attained $+19\%$
instead of $-38\%$ and for the softer $b$-jet $-16\%$ to be compared
with $-53\%$ for the fixed scale choice. The same pattern can be
noticed for the $p_T$ differential cross section of the hardest (the
softer) lepton. In details, for the hardest one we have obtained
$+8\%$ as a substitute to $-43\%$ whereas for the softer charged
lepton $-30\%$ rather than $-76\%$.

To summarise this part, the validity of the proposed dynamical scale
$\mu_0=H_T/4$, that is blind to the underlining top quark resonance
history, is confirmed. The size of NLO QCD corrections to all
presented observables has been reduced. Moreover, this judicious
choice of the scale has allowed us to obtain nearly constant ${\cal
K}$-factors in all dimensionful distributions that we have studied.

%
\subsection{Theoretical uncertainties for differential 
cross sections}
%

As a final step we have examined theoretical uncertainties for
differential cross sections for the dynamical scale choice
$\mu_0=H_T/4$.  The $\mu_0=m_t/2$ scale choice has proved to be
inadequate for the modelling of various differential distributions and
more importantly for the estimation of their theoretical errors in the
high $p_T$ regions. The latter phase space regions are simply not very
sensitive to the threshold contributions for the $t\bar{t}\gamma$
production that are well described by the fixed scale choice. For each
considered observable we have observed reduced theoretical errors as
compared to the $\mu_0=m_t/2$ scale choice. The effect is more
pronounce in the case of dimensionful observables in the high $p_T$
regions. Thus, for example we can see from
Fig.~\ref{fig:distribution1_ht} where the differential cross section
as a function of the transverse momentum of the hard photon is
plotted, that theoretical error is now up to  $\pm 8\%$ at
NLO (up to $\pm 36\%$ at LO) as compared to $\pm 22\%$ at NLO ($\pm
40\%$ at LO) for $\mu_0=m_t/2$. When considering $p_T$ distribution of
the hardest and the softer $b$-jet, depicted in
Fig.~\ref{fig:distribution3_ht}, the theoretical error at NLO is
reduced from $\pm 47\%$ and $\pm 78\%$ down  to $\pm 10\%$ and $\pm
 18\%$ respectively. The most dramatic effect can be seen in the case
 of $p_T$ distribution of the hardest and the softer lepton, also
 given in Fig.~\ref{fig:distribution3_ht}. In that case instead of
 theoretical errors up to $\pm 56\%$ and $\pm 186\%$ we have received 
 theoretical errors up to only $\pm 7\%$ and $\pm 31\%$ respectively.
  
To recapitulate this part, the dynamical scale $\mu_0=H_T/4$, which
has been considered in our analysis, has proven to be very effective
in stabilising the perturbative convergence in the phase space regions
far away from the $2m_t$ threshold and in providing small theoretical
uncertainties as estimated by the scale variation. For all considered
observables the latter are below $10\%-30\%$.

%
\section{Summary and Outlook}
\label{summary}
%
 
We have presented the first complete higher order predictions for the
$pp\to t\bar{t}\gamma$ process in the di-lepton channel for the LHC
run II energy of $\sqrt{s}=13$ TeV. With our inclusive cuts and for
$\mu_R=\mu_F=\mu_0=m_t/2$, NLO predictions reduced the unphysical
scale dependence by a factor of $2.5$ and lowered the total rate by
about $10\%$ compared to LO predictions. The theoretical
uncertainty of the NLO cross  section as obtained from the scale
dependence has been estimated to be $14\%$. By comparison the PDF
uncertainties are negligible at the level of $4\%$ only.  On the other hand, for
$\mu_R=\mu_F=\mu_0=H_T/4$ the full $pp$ cross section has received
positive and small NLO QCD corrections of $2.5\%$. Additionally, the
inclusion of higher order effects has reduced the theoretical error by
a factor of $5.5$. Specifically, the theoretical
uncertainties due to scale dependence are now at the $6\%$ level only,
however,  they are still larger than the PDF uncertainties. 

Even though NLO QCD corrections to the total cross section vary from
moderate to small depending on the scale choice their impact on
differential distributions is much larger.  Independently of the scale
choice for some dimensionless observables shape distortions of more
than $100\%$ have been observed. The prominent example comprises the
differential cross section as a function of the separation in the
rapidity-azimuthal angle plane between the hard photon and the softer
$b$-jet, $\Delta R_{b_2,\gamma}$. In the case of this observable,
which is relevant for new physics searches, shape distortions up to
$150\%$ ($128\%$) have been obtained for $\mu_0=m_t/2$
($\mu_0=H_T/4$). For the dimensionful observables presented in this
paper, however, the dynamical scale choice has helped to obtain almost
flat differential ${\cal K}$-factors as well as to stabilise the high
$p_T$ regions, which are very poorly described by NLO results with the
fixed scale choice.  Also in the case of differential
observables the PDF uncertainties have been examined. Similarly to the
total cross section case their size is negligible when comparing to scale
uncertainties. We repeat at this point  that additional
theoretical effects should be investigated. Among others the
size of NLO electroweak effects has to be calculated for the $pp \to
e^+\nu_e \mu^- \bar{\nu}_\mu b\bar{b}\gamma$ cross section and for
various differential cross sections. We plan to include such effects
in a future publication.

In addition, the size of
the top quark off-shell effects for the total cross section has been
estimated to be $\lesssim 2.5\%$. Their influence on differential
distributions and extraction of the SM parameters, however, might be
much stronger, as has already been shown in case of $t\bar{t}$ and
$t\bar{t}j$ production \cite{AlcarazMaestre:2012vp,Heinrich:2017bqp,
Bevilacqua:2017ipv}. Again, we leave such studies for the
future.

Our theoretical predictions are stored in the form of the Ntuples
files  and are available upon request. Specifically, they
are stored in the form of modified Les Houches event files and ROOT
files, that might be directly employed in experimental studies at the
LHC. They can be used for example to change kinematical cuts or to
define new observables. The latter can be obtained without need of any
additional rerunning of the code. Moreover, any change in the
renormalisation or factorisation scale choice or in the PDF set can be
accommodated by simple reweighting of these files. Thus, they can be
employed to study broad phenomenological aspects of top quark physics
at the LHC.

%
\section*{Acknowledgement}
%

The work of M.W. and T.W.  was supported by
the DFG Research Grant {\it "Top-Quarks under the LHCs Magnifying
Glass: From Process Modelling to Parameter Extraction''}. Furthermore,
the work of H.B.H. was supported by a Rutherford Grant
ST/M004104/1. The research of G.B. was supported by grant K 125105 of
the National Research, Development and Innovation Office in
Hungary. Simulations were performed with computing resources granted
by RWTH Aachen University under project {\tt rwth0211}.



\begin{thebibliography}{99}

\bibitem{Fael:2013ira}
  M.~Fael and T.~Gehrmann,
  {\it Probing top quark electromagnetic dipole moments in
    single-top-plus-photon production},
  Phys.\ Rev.\ D {\bf 88} (2013) 033003
 \href{https://arxiv.org/abs/1307.1349}{\tt [arXiv:1307.1349 [hep-ph]]}.

\bibitem{Saavedra:2014vta}
  J.~A.~Aguilar-Saavedra, E.~Alvarez, A.~Juste and F.~Rubbo,
  {\it Shedding light on the $t \bar t$ asymmetry: the photon handle},
  JHEP {\bf 1404} (2014) 188
 \href{https://arxiv.org/abs/1402.3598}{\tt  [arXiv:1402.3598 [hep-ph]]}.

\bibitem{Schulze:2016qas}
  M.~Schulze and Y.~Soreq,
  {\it Pinning down electroweak dipole operators of the top quark},
  Eur.\ Phys.\ J.\ C {\bf 76} (2016) no.8,  466
  \href{https://arxiv.org/abs/1603.08911}{\tt  [arXiv:1603.08911 [hep-ph]]}.

\bibitem{Etesami:2016rwu}
  S.~M.~Etesami, S.~Khatibi and M.~Mohammadi Najafabadi,
  {\it Measuring anomalous  $WW\gamma $ and  $t\bar{t}\gamma $
    couplings using top+$\gamma $ production at the LHC},
  Eur.\ Phys.\ J.\ C {\bf 76} (2016) no.10,  533
 \href{https://arxiv.org/abs/1606.02178}{\tt  [arXiv:1606.02178 [hep-ph]]}.

\bibitem{Baur:2001si}
  U.~Baur, M.~Buice and L.~H.~Orr,
  {\it Direct measurement of the top quark charge at hadron colliders},
    Phys.\ Rev.\ D {\bf 64} (2001) 094019 
  \href{http://arxiv.org/abs/hep-ph/0106341}{\tt [hep-ph/0106341]}

\bibitem{Baur:2004uw}
  U.~Baur, A.~Juste, L.~H.~Orr and D.~Rainwater,
  {\it Probing electroweak top quark couplings at hadron colliders},
   Phys.\ Rev.\ D {\bf 71} (2005) 054013
   \href{http://arxiv.org/abs/hep-ph/0412021}{\tt [hep-ph/0412021]}

\bibitem{Aaltonen:2011sp}
  T.~Aaltonen {\it et al.} [CDF Collaboration],
  {\it Evidence for $t\bar{t}\gamma$ Production and Measurement of
    $\sigma_{t\bar{t}\gamma} / \sigma_{t\bar{t}}$},
  Phys.\ Rev.\ D {\bf 84} (2011) 031104
   \href{http://arxiv.org/abs/1106.3970}{\tt [arXiv:1106.3970 [hep-ex]]}.

\bibitem{Aad:2015uwa}
  G.~Aad {\it et al.} [ATLAS Collaboration],
  {\it Observation of top-quark pair production in association with a
    photon and measurement of the $t\bar{t}\gamma$ production cross
    section in pp collisions at $\sqrt{s}=7$ TeV using the ATLAS
    detector},
  Phys.\ Rev.\ D {\bf 91} (2015) no.7,  072007
   \href{http://arxiv.org/abs/1502.00586}{\tt [arXiv:1502.00586 [hep-ex]]}.

\bibitem{Aaboud:2017era}
  M.~Aaboud {\it et al.} [ATLAS Collaboration],
  {\it Measurement of the $ t\bar{t} \gamma $ production cross
    section in proton-proton collisions at $ \sqrt{s}=8 $ TeV with the
    ATLAS detector},
  JHEP {\bf 1711} (2017) 086
   \href{http://arxiv.org/abs/1706.03046}{\tt [arXiv:1706.03046 [hep-ex]]}.

\bibitem{Sirunyan:2017iyh}
  A.~M.~Sirunyan {\it et al.} [CMS Collaboration],
  {\it Measurement of the semi-leptonic $t\bar{t}+\gamma$ production
    cross section in   pp collisions at $ \sqrt{s}=8 $ TeV},
  JHEP {\bf 1710} (2017) 006
  \href{http://arxiv.org/abs/1706.08128}{\tt [arXiv:1706.08128 [hep-ex]]}.

\bibitem{PengFei:2009ph}
  P.~F.~Duan, W.~G.~Ma, R.~Y.~Zhang, L.~Han, L.~Guo and S.~M.~Wang,
  {\it QCD corrections to associated production of $t\bar t\gamma$ at
    hadron colliders},
  Phys.\ Rev.\ D {\bf 80} (2009) 014022
    \href{http://arxiv.org/abs/0907.1324}{\tt [arXiv:0907.1324 [hep-ph]]}.

\bibitem{PengFei:2011qg}
  D.~Peng-Fei, Z.~Ren-You, M.~Wen-Gan, H.~Liang, G.~Lei and W.~Shao-Ming,
  {\it Next-to-leading order QCD corrections to $t\bar t\gamma$
    production at the 7 TeV LHC}, 
  Chin.\ Phys.\ Lett.\  {\bf 28} (2011) 111401
    \href{http://arxiv.org/abs/1110.2315}{\tt [arXiv:1110.2315 [hep-ph]]}.

\bibitem{Maltoni:2015ena}
  F.~Maltoni, D.~Pagani and I.~Tsinikos,
  {\it Associated production of a top-quark pair with vector bosons at
    NLO in QCD: impact on $t\bar{t}H$ searches at the LHC},
  JHEP {\bf 1602} (2016) 113
   \href{http://arxiv.org/abs/1507.05640}{\tt [arXiv:1507.05640 [hep-ph]]}.

\bibitem{Duan:2016qlc}
  P.~F.~Duan, Y.~Zhang, Y.~Wang, M.~Song and G.~Li,
  {\it Electroweak corrections to top quark pair production in
    association with a hard photon at hadron colliders},
  Phys.\ Lett.\ B {\bf 766} (2017) 102
  \href{http://arxiv.org/abs/1612.00248}{\tt [arXiv:1612.00248 [hep-ph]]}.

\bibitem{Nason:2004rx}
  P.~Nason,
  {\it A New method for combining NLO QCD with shower Monte Carlo
    algorithms},
  JHEP {\bf 0411} (2004) 040
  \href{http://arxiv.org/abs/hep-ph/0409146}{\tt [hep-ph/0409146]}.
 
\bibitem{Frixione:2007vw}
  S.~Frixione, P.~Nason and C.~Oleari,
  {\it Matching NLO QCD computations with Parton Shower simulations:
    the POWHEG method}, JHEP {\bf 0711} (2007) 070 
 \href{http://arxiv.org/abs/0709.2092}{\tt [arXiv:0709.2092 [hep-ph]]}.

\bibitem{Kardos:2014zba}
  A.~Kardos and Z.~Trocsanyi,
  {\it Hadroproduction of t anti-t pair in association with an
    isolated photon at NLO accuracy matched with parton shower},
  JHEP {\bf 1505} (2015) 090
   \href{http://arxiv.org/abs/1406.2324}{\tt [arXiv:1406.2324 [hep-ph]]}.

\bibitem{Melnikov:2011ta}
  K.~Melnikov, M.~Schulze and A.~Scharf,
  {\it QCD corrections to top quark pair production in association
    with a photon at hadron colliders}, 
  Phys.\ Rev.\ D {\bf 83} (2011) 074013
     \href{http://arxiv.org/abs/1102.1967}{\tt [arXiv:1102.1967 [hep-ph]]}.

\bibitem{Bern:2013zja}
  Z.~Bern, L.~J.~Dixon, F.~Febres Cordero, S.~Höche, H.~Ita,
  D.~A.~Kosower and D.~Maitre,
  {\it Ntuples for NLO Events at Hadron Colliders}, 
  Comput.\ Phys.\ Commun.\  {\bf 185} (2014) 1443
 \href{http://arxiv.org/abs/1310.7439}{\tt [arXiv:1310.7439 [hep-ph]]}.

\bibitem{Alwall:2006yp}
  J.~Alwall {\it et al.},
  {\it A Standard format for Les Houches event files},
  Comput.\ Phys.\ Commun.\  {\bf 176} (2007) 300
   \href{http://arxiv.org/abs/hep-ph/0609017}{\tt  [hep-ph/0609017]}.

\bibitem{Antcheva:2009zz}
  I.~Antcheva {\it et al.},
  {\it ROOT: A C++ framework for petabyte data storage, statistical
    analysis and visualization},
  Comput.\ Phys.\ Commun.\  {\bf 180} (2009) 2499
 \href{https://arxiv.org/abs/1508.07749}{\tt  [arXiv:1508.07749 [physics.data-an]]}.

\bibitem{Denner:2010jp}
  A.~Denner, S.~Dittmaier, S.~Kallweit and S.~Pozzorini,
  {\it NLO QCD corrections to WWbb production at hadron colliders},
  Phys.\ Rev.\ Lett.\  {\bf 106} (2011) 052001
  \href{http://arxiv.org/abs/1012.3975}{\tt [arXiv:1012.3975 [hep-ph]]}.

\bibitem{Bevilacqua:2010qb}
  G.~Bevilacqua, M.~Czakon, A.~van Hameren, C.~G.~Papadopoulos and
  M.~Worek,
  {\it Complete off-shell effects in top quark pair hadroproduction
    with leptonic decay at next-to-leading order},
  JHEP {\bf 1102} (2011) 083
    \href{http://arxiv.org/abs/1012.4230}{\tt [arXiv:1012.4230 [hep-ph]]}.

\bibitem{Denner:2012yc}
  A.~Denner, S.~Dittmaier, S.~Kallweit and S.~Pozzorini,
  {\it NLO QCD corrections to off-shell top-antitop production with
    leptonic decays at hadron colliders},
  JHEP {\bf 1210} (2012) 110
   \href{http://arxiv.org/abs/1207.5018}{\tt [arXiv:1207.5018 [hep-ph]]}.

\bibitem{Frederix:2013gra}
  R.~Frederix,
  {\it Top Quark Induced Backgrounds to Higgs Production in the
    $WW^{(*)}\to ll\nu\nu$ Decay Channel at Next-to-Leading-Order in
    QCD},
  Phys.\ Rev.\ Lett.\  {\bf 112} (2014) no.8,  082002
  \href{http://arxiv.org/abs/1311.4893}{\tt [arXiv:1311.4893 [hep-ph]]}.

\bibitem{Heinrich:2013qaa}
  G.~Heinrich, A.~Maier, R.~Nisius, J.~Schlenk and J.~Winter,
  {\it NLO QCD corrections to $W^{+} W^{-}b\bar{b}$ production with
    leptonic decays in the light of top quark mass and asymmetry
    measurements},
  JHEP {\bf 1406} (2014) 158
    \href{http://arxiv.org/abs/1312.6659}{\tt [arXiv:1312.6659 [hep-ph]]}.

\bibitem{Denner:2017kzu}
  A.~Denner and M.~Pellen,
  {\it Off-shell production of top-antitop pairs in the lepton+jets
    channel at NLO QCD},
  JHEP {\bf 1802} (2018) 013
  \href{http://arxiv.org/abs/1711.10359}{\tt [arXiv:1711.10359 [hep-ph]]}.

\bibitem{Denner:2015yca}
  A.~Denner and R.~Feger,
  {\it NLO QCD corrections to off-shell top-antitop production with
    leptonic decays in association with a Higgs boson at the LHC},
  JHEP {\bf 1511} (2015) 209
    \href{http://arxiv.org/abs/1506.07448}{\tt [arXiv:1506.07448 [hep-ph]]}.

\bibitem{Bevilacqua:2015qha}
  G.~Bevilacqua, H.~B.~Hartanto, M.~Kraus and M.~Worek,
  {\it Top Quark Pair Production in Association with a Jet with
    Next-to-Leading-Order QCD Off-Shell Effects at the Large Hadron
    Collider},
  Phys.\ Rev.\ Lett.\  {\bf 116} (2016) no.5,  052003
    \href{http://arxiv.org/abs/1509.09242}{\tt [arXiv:1509.09242 [hep-ph]]}.

\bibitem{Bevilacqua:2016jfk}
  G.~Bevilacqua, H.~B.~Hartanto, M.~Kraus and M.~Worek,
  {\it Off-shell Top Quarks with One Jet at the LHC: A comprehensive
    analysis at NLO QCD},
  JHEP {\bf 1611} (2016) 098
    \href{http://arxiv.org/abs/1609.01659}{\tt [arXiv:1609.01659 [hep-ph]]}.

\bibitem{Denner:2016jyo}
  A.~Denner and M.~Pellen,
  {\it NLO electroweak corrections to off-shell top-antitop production
    with leptonic decays at the LHC},
  JHEP {\bf 1608} (2016) 155
   \href{http://arxiv.org/abs/1607.05571}{\tt [arXiv:1607.05571 [hep-ph]]}.

\bibitem{Denner:2016wet}
  A.~Denner, J.~N.~Lang, M.~Pellen and S.~Uccirati,
  {\it Higgs production in association with off-shell top-antitop
    pairs at NLO EW and QCD at the LHC},
  JHEP {\bf 1702} (2017) 053
  \href{http://arxiv.org/abs/1612.07138}{\tt [arXiv:1612.07138 [hep-ph]]}.

\bibitem{Papadopoulos:2005ky}
  C.~G.~Papadopoulos and M.~Worek,
  {\it Multi-parton cross sections at hadron colliders},
  Eur.\ Phys.\ J.\ C {\bf 50} (2007) 843
   \href{http://arxiv.org/abs/hep-ph/hep-ph/0512150}{\tt  [hep-ph/0512150]}.

\bibitem{Czakon:2009ss}
  M.~Czakon, C.~G.~Papadopoulos and M.~Worek,
  {\it Polarizing the Dipoles},
  JHEP {\bf 0908} (2009) 085
   \href{http://arxiv.org/abs/0905.0883}{\tt [arXiv:0905.0883 [hep-ph]]}.

\bibitem{Cafarella:2007pc}
  A.~Cafarella, C.~G.~Papadopoulos and M.~Worek,
  {\it Helac-Phegas: A Generator for all parton level processes}, 
  Comput.\ Phys.\ Commun.\  {\bf 180} (2009) 1941
   \href{http://arxiv.org/abs/0710.2427}{\tt [arXiv:0710.2427 [hep-ph]]}.

\bibitem{Papadopoulos:2000tt}
  C.~G.~Papadopoulos,
  {\it PHEGAS: A Phase space generator for automatic cross-section
    computation},
  Comput.\ Phys.\ Commun.\  {\bf 137} (2001) 247
 \href{http://arxiv.org/abs/hep-ph/0007335}{\tt  [hep-ph/0007335]}.

\bibitem{vanHameren:2007pt}
  A.~van Hameren,
  {\it PARNI for importance sampling and density estimation},
  Acta Phys.\ Polon.\ B {\bf 40} (2009) 259
 \href{http://arxiv.org/abs/0710.2448}{\tt [arXiv:0710.2448 [hep-ph]]}.

\bibitem{vanHameren:2010gg}
  A.~van Hameren,
  {\it Kaleu: A General-Purpose Parton-Level Phase Space Generator},
  \href{http://arxiv.org/abs/1003.4953}{\tt [arXiv:1003.4953 [hep-ph]]}.

\bibitem{Nogueira:1991ex}
  P.~Nogueira,
  {\it Automatic Feynman graph generation},
  J.\ Comput.\ Phys.\  {\bf 105} (1993) 279.

\bibitem{Badger:2010nx}
  S.~Badger, B.~Biedermann and P.~Uwer,
  {\it NGluon: A Package to Calculate One-loop Multi-gluon Amplitudes},
  Comput.\ Phys.\ Commun.\  {\bf 182} (2011) 1674
  \href{http://arxiv.org/abs/1011.2900}{\tt    [arXiv:1011.2900 [hep-ph]]}.

\bibitem{vanHameren:2009dr}
  A.~van Hameren, C.~G.~Papadopoulos and R.~Pittau,
  {\it Automated one-loop calculations: A Proof of concept},
  JHEP {\bf 0909} (2009) 106
   \href{http://arxiv.org/abs/0903.4665}{\tt [arXiv:0903.4665 [hep-ph]]}.

\bibitem{Ossola:2007ax}
  G.~Ossola, C.~G.~Papadopoulos and R.~Pittau,
  {\it CutTools: A Program implementing the OPP reduction method to
    compute one-loop amplitudes},
  JHEP {\bf 0803} (2008) 042
    \href{http://arxiv.org/abs/0711.3596}{\tt [arXiv:0711.3596 [hep-ph]]}.

\bibitem{Bevilacqua:2011xh}
  G.~Bevilacqua, M.~Czakon, M.~V.~Garzelli, A.~van Hameren, A.~Kardos,
  C.~G.~Papadopoulos, R.~Pittau and M.~Worek,
  {\it HELAC-NLO},
  Comput.\ Phys.\ Commun.\  {\bf 184} (2013) 986
   \href{http://arxiv.org/abs/1110.1499}{\tt [arXiv:1110.1499 [hep-ph]]}.

\bibitem{Ossola:2006us}
  G.~Ossola, C.~G.~Papadopoulos and R.~Pittau,
  {\it Reducing full one-loop amplitudes to scalar integrals at the integrand level},
  Nucl.\ Phys.\ B {\bf 763} (2007) 147
   \href{http://arxiv.org/abs/hep-ph/0609007}{\tt   [hep-ph/0609007]}.

\bibitem{Ossola:2008xq}
  G.~Ossola, C.~G.~Papadopoulos and R.~Pittau,
  {\it On the Rational Terms of the one-loop amplitudes},
  JHEP {\bf 0805} (2008) 004
  \href{http://arxiv.org/abs/0802.1876}{\tt   [arXiv:0802.1876 [hep-ph]]}.

\bibitem{Draggiotis:2009yb}
  P.~Draggiotis, M.~V.~Garzelli, C.~G.~Papadopoulos and R.~Pittau,
  {\it Feynman Rules for the Rational Part of the QCD 1-loop amplitudes},
  JHEP {\bf 0904} (2009) 072
  \href{http://arxiv.org/abs/0903.0356}{\tt   [arXiv:0903.0356 [hep-ph]]}.

\bibitem{Denner:1999gp}
  A.~Denner, S.~Dittmaier, M.~Roth and D.~Wackeroth,
  {\it Predictions for all processes  $e^+e^- \to 4$ fermions + $\gamma$},
  Nucl.\ Phys.\ B {\bf 560} (1999) 33
  \href{http://arxiv.org/abs/hep-ph/9904472}{\tt [hep-ph/9904472]}.

\bibitem{Denner:2005fg}
  A.~Denner, S.~Dittmaier, M.~Roth and L.~H.~Wieders,
  {\it Electroweak corrections to charged-current $e^+ e^- \to 4$ fermion
    processes: Technical details and further results},
  Nucl.\ Phys.\ B {\bf 724} (2005) 247,
   Erratum: [Nucl.\ Phys.\ B {\bf 854} (2012) 504]
     \href{http://arxiv.org/abs/hep-ph/0505042}{\tt [hep-ph/0505042]}.

\bibitem{vanHameren:2010cp}
  A.~van Hameren,
  {\it OneLOop: For the evaluation of one-loop scalar functions},
  Comput.\ Phys.\ Commun.\  {\bf 182} (2011) 2427
    \href{http://arxiv.org/abs/1007.4716}{\tt [arXiv:1007.4716 [hep-ph]]}.

\bibitem{Catani:1996vz}
  S.~Catani and M.~H.~Seymour,
  {\it A General algorithm for calculating jet cross-sections in NLO QCD},
  Nucl.\ Phys.\ B {\bf 485} (1997) 291,
   Erratum: [Nucl.\ Phys.\ B {\bf 510} (1998) 503]
  \href{http://arxiv.org/abs/hep-ph/9605323}{\tt [hep-ph/9605323]}.

\bibitem{Catani:2002hc}
  S.~Catani, S.~Dittmaier, M.~H.~Seymour and Z.~Trocsanyi,
  {\it The Dipole formalism for next-to-leading order QCD calculations
    with massive partons},
  Nucl.\ Phys.\ B {\bf 627} (2002) 189
   \href{http://arxiv.org/abs/hep-ph/0201036}{\tt [hep-ph/0201036]}.

\bibitem{Bevilacqua:2013iha}
  G.~Bevilacqua, M.~Czakon, M.~Kubocz and M.~Worek,
  {\it Complete Nagy-Soper subtraction for next-to-leading order
    calculations in QCD},
  JHEP {\bf 1310} (2013) 204
    \href{http://arxiv.org/abs/1308.5605}{\tt [arXiv:1308.5605 [hep-ph]]}.

\bibitem{Frixione:1995ms}
  S.~Frixione, Z.~Kunszt and A.~Signer,
  {\it Three jet cross-sections to next-to-leading order}, 
  Nucl.\ Phys.\ B {\bf 467} (1996) 399
  \href{https://arxiv.org/abs/hep-ph/9512328}{\tt [hep-ph/9512328]}.

\bibitem{Nagy:1998bb}
  Z.~Nagy and Z.~Trocsanyi,
  {\it Next-to-leading order calculation of four jet observables in
    electron positron annihilation},
  Phys.\ Rev.\ D {\bf 59} (1999) 014020
   Erratum: [Phys.\ Rev.\ D {\bf 62} (2000) 099902]
 \href{https://arxiv.org/abs/hep-ph/9806317}{\tt [hep-ph/9806317]}.

\bibitem{Nagy:2003tz}
  Z.~Nagy,
  {\it Next-to-leading order calculation of three jet observables in
    hadron hadron collision},
  Phys.\ Rev.\ D {\bf 68} (2003) 094002
  \href{https://arxiv.org/abs/hep-ph/0307268}{\tt [hep-ph/0307268]}.

\bibitem{Campbell:2005bb}
  J.~M.~Campbell and F.~Tramontano,
  {\it Next-to-leading order corrections to Wt production and decay},
  Nucl.\ Phys.\ B {\bf 726} (2005) 109
  \href{https://arxiv.org/abs/hep-ph/0506289}{\tt [hep-ph/0506289]}.

\bibitem{Bevilacqua:2009zn}
  G.~Bevilacqua, M.~Czakon, C.~G.~Papadopoulos, R.~Pittau and M.~Worek,
  {\it Assault on the NLO Wishlist: $pp \to t\bar{t} b\bar{b}$},'
  JHEP {\bf 0909} (2009) 109
  \href{https://arxiv.org/abs/0907.4723}{\tt [arXiv:0907.4723 [hep-ph]]}.

\bibitem{Czakon:2015cla}
  M.~Czakon, H.~B.~Hartanto, M.~Kraus and M.~Worek,
  {\it Matching the Nagy-Soper parton shower at next-to-leading order},
  JHEP {\bf 1506} (2015) 033
  \href{https://arxiv.org/abs/1502.00925}{\tt [arXiv:1502.00925 [hep-ph]]}.

\bibitem{Jezabek:1988iv}
  M.~Je\.zabek and J.~H.~K\"uhn, 
  {\it QCD Corrections to Semileptonic Decays of Heavy Quarks},
  Nucl.\ Phys.\ B {\bf 314} (1989) 1.

\bibitem{Czakon:2011xx}
  M.~Czakon and A.~Mitov,
  {\it Top++: A Program for the Calculation of the Top-Pair
    Cross-Section at Hadron Colliders},
  Comput.\ Phys.\ Commun.\  {\bf 185} (2014) 2930
   \href{http://arxiv.org/abs/1112.5675}{\tt [arXiv:1112.5675 [hep-ph]]}.

\bibitem{Dulat:2015mca}
  S.~Dulat {\it et al.},
  {\it New parton distribution functions from a global analysis of
    quantum chromodynamics'},
   Phys.\ Rev.\ D {\bf 93} (2016) no.3,  033006
  \href{http://arxiv.org/abs/1506.07443}{\tt [arXiv:1506.07443 [hep-ph]]}.

\bibitem{Cacciari:2008gp}
  M.~Cacciari, G.~P.~Salam and G.~Soyez,
  {\it The Anti-k(t) jet clustering algorithm},
  JHEP {\bf 0804} (2008) 063
   \href{http://arxiv.org/abs/0802.1189}{\tt [arXiv:0802.1189 [hep-ph]]}.

\bibitem{Frixione:1998jh}
  S.~Frixione,
 {\it Isolated photons in perturbative QCD},
  Phys.\ Lett.\ B {\bf 429} (1998) 369
  \href{http://arxiv.org/abs/hep-ph/9801442}{\tt [hep-ph/9801442]}.

\bibitem{Butterworth:2015oua}
  J.~Butterworth {\it et al.},
  {\it DF4LHC recommendations for LHC Run II},
  J.\ Phys.\ G {\bf 43} (2016) 023001
  \href{http://arxiv.org/abs/1510.03865}{\tt   [arXiv:1510.03865 [hep-ph]]}.

\bibitem{Harland-Lang:2014zoa}
  L.~A.~Harland-Lang, A.~D.~Martin, P.~Motylinski and R.~S.~Thorne,
  {\it Parton distributions in the LHC era: MMHT 2014 PDFs},
  Eur.\ Phys.\ J.\ C {\bf 75} (2015) no.5,  204
   \href{http://arxiv.org/abs/1412.3989}{\tt [arXiv:1412.3989 [hep-ph]]}.

\bibitem{Ball:2014uwa}
  R.~D.~Ball {\it et al.} [NNPDF Collaboration],
  {\it Parton distributions for the LHC Run II}, 
  JHEP {\bf 1504} (2015) 040
  \href{http://arxiv.org/abs/1410.8849}{\tt [arXiv:1410.8849 [hep-ph]]}.

\bibitem{AlcarazMaestre:2012vp}
  J.~Alcaraz Maestre {\it et al.} [SM and NLO Multileg Working Group
  and SM MC Working Group],
  {\it The SM and NLO Multileg and SM MC Working Groups: Summary Report},
  \href{http://arxiv.org/abs/1203.6803}{\tt [arXiv:1203.6803 [hep-ph]]}.

\bibitem{Heinrich:2017bqp}
  G.~Heinrich, A.~Maier, R.~Nisius, J.~Schlenk, M.~Schulze, L.~Scyboz and J.~Winter,
  {\it NLO and off-shell effects in top quark mass determinations},
  \href{http://arxiv.org/abs/1709.08615}{\tt [arXiv:1709.08615 [hep-ph]]}.

\bibitem{Bevilacqua:2017ipv}
  G.~Bevilacqua, H.~B.~Hartanto, M.~Kraus, M.~Schulze and M.~Worek,
  {\it Top quark mass studies with $t\bar{t}j$ at the LHC},
  \href{http://arxiv.org/abs/1710.07515}{\tt [arXiv:1710.07515 [hep-ph]]}.

\end{thebibliography}
\end{document}